\documentclass[%
 reprint,
superscriptaddress,
 nobibnotes,
 amsmath,amssymb,
 aps,
prb,
]{revtex4-1}

\usepackage{graphicx}% Include figure files
\usepackage{dcolumn}% Align table columns on decimal point
\usepackage{bm}% bold math
\usepackage{xcolor,soul} %color text
\usepackage{braket} % Dirac braket
\usepackage{float}
\usepackage{soul}
\usepackage{footnote}
\usepackage[symbol]{footmisc}

\DefineFNsymbols{mySymbols}{{\ensuremath\dagger}{*}{\ensuremath\ddagger}{\ensuremath{\dagger\dagger}}{\ensuremath{\ddagger\ddagger}}}
\setfnsymbol{mySymbols}

\begin{document}
\title{Phonon softening near topological phase transitions}
\author{Shengying Yue}
\thanks{These authors contributed equally to this work}
\affiliation{Department of Mechanical Engineering, University of California, Santa Barbara, CA 93106, USA}
\author{Bowen Deng}
\thanks{These authors contributed equally to this work}
\affiliation{Department of Mechanical Engineering, University of California, Santa Barbara, CA 93106, USA}
\author{Yanming Liu}
\thanks{These authors contributed equally to this work}
\affiliation{Department of Mechanical Engineering, University of California, Santa Barbara, CA 93106, USA}
\affiliation{School of Aerospace Engineering, Tsinghua University, Beijing, 100084, China}
\author{Yujie Quan}
\author{Runqing Yang}
\author{Bolin Liao}\email{bliao@ucsb.edu}
\affiliation{Department of Mechanical Engineering, University of California, Santa Barbara, CA 93106, USA}
\date{\today}

\begin{abstract}
Topological phase transitions occur when the electronic bands change their topological properties, typically featuring the closing of the band gap. While the influence of topological phase transitions on electronic and optical properties has been extensively studied, its implication on phononic properties and thermal transport remain unexplored. In this work, we use first-principles simulations to show that certain phonon modes are significantly softened near topological phase transitions, leading to increased phonon-phonon scattering and reduced lattice thermal conductivity. We demonstrate this effect using two model systems: pressure induced topological phase transition in $\rm ZrTe_5$ and chemical composition induced topological phase transition in $\rm{Hg_{1-x}Cd_{x}Te}$. We attribute the phonon softening to emergent Kohn anomalies associated with the closing of the band gap. Our study reveals the strong connection between electronic band structures and lattice instabilities, and opens up a potential direction towards controlling heat conduction in solids.  
\end{abstract}


\maketitle

\section{Introduction}
{\color{red}}
Topological materials possess electronic bands with nontrivial topological indices\cite{yan2012topological}. Depending on whether a bulk band gap exists, topological materials can be classified into topological insulators with a finite band gap and topological semimetals without a band gap. Due to the distinct topological indices in different insulating states (e.g. normal insulator, strong topological insulator and weak topological insulator states), direct transitions between these states cannot be achieved by continuous tuning of material parameters and/or external conditions\cite{mutch2019evidence,orlita2014observation,teppe2016temperature}. Instead, these transitions must go through a semimetal phase associated with the closing of the band gap. Recently, these so-called topological phase transitions (TPTs) have attracted significant research efforts. TPTs are not only of fundamental scientific interest, but also signal potential practical methods to drastically alter material properties with small changes of external parameters. 

Previous studies have predicted or demonstrated TPTs induced in different ways, e.g. by photoexcitation\cite{ezawa2013photoinduced} or applying an electric field\cite{collins2018electric}, or by changing the strain\cite{zhu2012topological,mutch2019evidence,fan2017transition}, the quantum well thickness\cite{bernevig2006quantum}, the chemical composition\cite{orlita2014observation,xu2011topological,wu2013sudden,dziawa2012topological,xu2012observation}, temperature\cite{xu2018temperature} and pressure\cite{chen2019enhancement}. Most of the previous studies focused on the electrical transport and optical properties across the TPTs. Some of these interesting properties are enabled by the emergent topological semimetal phases, such as the Dirac fermions with ultrahigh mobilities\citep{wang2013-Cd3As2,Ominato2014,orlita2014observation,neupane2014observation,schumann2016molecular}, the Fermi arc surface states\citep{Wan2011-Fermi-arc,Xu2015-Fermi-arc,moll2016transport} and highly temperature-sensitive optical properties\citep{chorsi2019widely}. On the other hand, the phonon properties and phonon-mediated thermal conduction across TPTs have not been examined in detail. In a recent study\cite{yue2019soft}, we reported the existence of ultrasoft optical phonons in the topological Dirac semimetal cadmium arsenide ($\rm{Cd_3As_2}$), which gives rise to strong phonon-phonon scatterings and a low lattice thermal conductivity. The phonon softening was attributed to potentially strong Kohn anomalies\cite{Kohn1959} associated with the Dirac nodes. In parallel, Nguyen et al. reported experimental observation of Kohn anomalies in a topological Weyl semimetal tantalum phosphide (TaP)\cite{nguyen2020topological}. Therefore, it is interesting to see how the phonon properties continuously evolve across the TPTs and whether the TPTs offer a promising route towards sensitive control of the thermal conduction.

In this work, we use first-principles simulations to examine the phonon properties across TPTs in two model systems: pressure induced TPT in zirconium pentatelluride ($\mathrm{ZrTe_5}$)\citep{weng2014,mutch2019evidence,zhou2016pressure} and chemical composition induced TPT in mercury cadmium telluride  ($\mathrm{Hg_{1-x}Cd_{x}Te}$)\cite{orlita2014observation,teppe2016temperature}. We show that certain acoustic phonons and low-lying optical phonons significantly soften near the TPTs when the semimetal phases emerge, at locations in the Brillouin zone that match the characteristics of Kohn anomalies. We further demonstrate that the phonon softening largely increases the phase space for phonon-phonon scatterings and reduces the lattice thermal conductivity. Our findings point to the direct connection between the electronic band topology and lattice instabilities, potentially adding to the understanding of the coincidence between topological materials and good thermoelectric materials. Our results also suggest a novel route towards controlling solid-state thermal transport based on TPTs.   

\section{Computational Details}

\begin{figure}
\includegraphics[width=0.8\linewidth,clip]{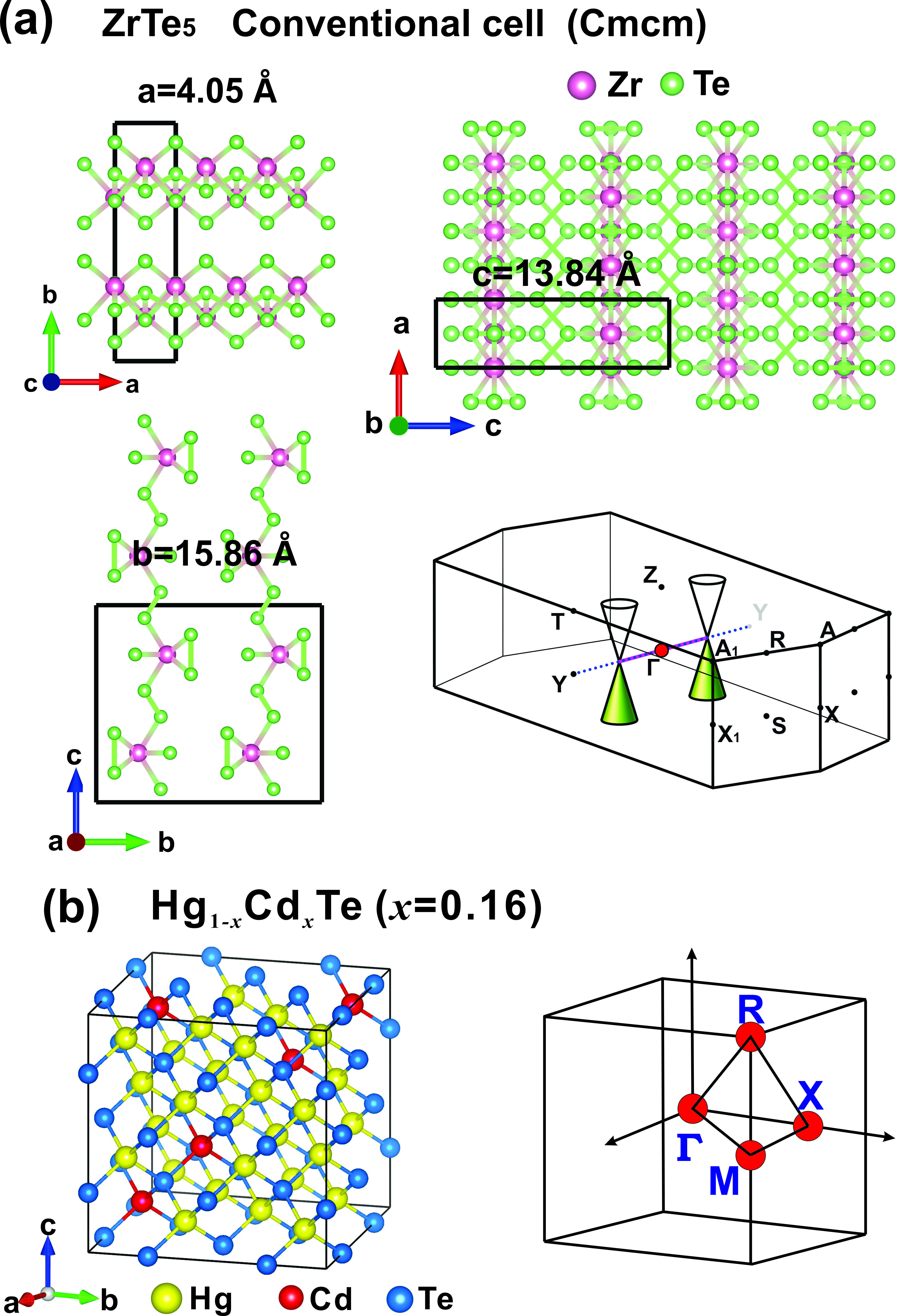}
\caption{\textbf{(a)} The crystal structure of $\rm ZrTe_5$ with calculated lattice constants and the corresponding first Brillouin zone, where the band touching points under 5 GPa pressure are labeled with two cones. \textbf{(b)} The crystal structure and the corresponding Brillouin zone of the $\rm Hg_{1-x}Cd_xTe$ alloy system.}
\label{fig:1}
\end{figure}

$\rm ZrTe_5$ is a van der Waals (vdW) layered material that crystallizes in the base-centered orthorhombic $Cmcm$ ($D^{17}_{2h}$) structure under ambient conditions\citep{zhou2016pressure}. In this structure, $\rm ZrTe_3$ chains run along the $a$ axis, which are connected by zigzag $\rm Te$ atoms to form two-dimensional (2D) sheets. These 2D sheets are further stacked along the $b$ axis. The crystal structure and the conventional cell with the corresponding Brillouin zone is shown in Fig.~\ref{fig:1}(a), where $\Gamma$-X, $\Gamma$-Y and $\Gamma$-Z are aligned with the $a$,$b$ and $c$ axis, respectively. Both $\mathrm{HgTe}$ and $\mathrm{CdTe}$ crystallize in the zinc-blende structure. The $\mathrm{Hg_{1-x}Cd_{x}Te}$ alloys with different Cd concentrations are simulated by randomly replacing Hg atoms with Cd atoms in supercells (Fig.~\ref{fig:1}(b)).

All of the structural optimizations, electronic and phonon properties were calculated based on the density functional theory (DFT) by employing the Vienna $ab~initio$ simulation package (VASP)\citep{vasp-01,vasp-02}. The projector augmented wave (PAW) method\citep{PAW-01,PAW-02} and generalized gradient approximation (GGA) with the Perdew-Burke-Ernzerhof (PBE) exchange-correlation functional were adpoted\citep{GGA}. For $\mathrm{ZrTe_5}$, the van der Waals (vdW) corrected optB86b-vdw functional\citep{vdw-01,vdw-02} was used throughout all calculations because $\mathrm{ZrTe_5}$ is a layered material for which the vdW correction is essential to obtain the correct inter-layer distance\citep{ZrTe5-01,ZrTe5-02}. The plane wave cutoff energy was set to $\mathrm{350~eV}$ for all materials. The Monkhorst-Pack $k$ meshes $\mathrm{9\times 9 \times 4}$ were taken for $\mathrm{ZrTe_5}$, and $\mathrm{8\times 8\times 8}$ for $\mathrm{Hg_{1-x}Cd_xTe}$. The $k$-mesh density and the energy cutoff of augmented plane waves were checked to ensure the convergence. Full optimizations were applied to both structures with the Hellmann-Feynman forces tolerance $\mathrm{0.0001 eV/\AA}$. The spin-orbit coupling (SOC) effect was included through all calculations.

To evaluate the phonon dispersions and lattice thermal conductivities, we further calculated the second and third order inter-atomic force constants (IFCs) using the finite-displacement approach\citep{phonopy}. In the calculations of the second and third order IFCs, we adopted $\rm 3\times 3\times 1$ supercells\citep{wang2018first} for $\rm ZrTe_5$ and $\rm 2\times 2\times 2$ supercells for $\mathrm{Hg_{1-x}Cd_xTe}$. The phonon dispersions based on the second order IFCs were calculated using the PHONOPY package\citep{phonopy}. The interactions between atoms were taken into account up to sixth nearest neighbors in third-order IFC calculations. We calculated the lattice thermal conductivity by solving the phonon Boltzmann transport equation (BTE) iteratively as implemented in ShengBTE\citep{shengbte-01,shengbte-02}. The $q$-mesh sampling for the BTE calculations of $\rm ZrTe_5$ and $\mathrm{Hg_{1-x}Cd_xTe}$ were $\rm 8\times 8\times 20$ and $\rm 10 \times 10 \times 10$, respectively. The convergence of the lattice thermal conductivity with respect to the $q$-mesh density and the interaction distance cutoff was checked. Our adoption of the finite-displacement approach implicitly determines that only the static, or adiabatic, Kohn anomalies can be captured\cite{piscanec2004kohn}, while the dynamic effect\cite{nguyen2020topological} requires explicit electron-phonon coupling calculations\cite{lazzeri2006nonadiabatic} which is computationally intractable for complex crystal structures such as that of $\rm ZrTe_5$. For simple crystal structures of $\rm HgTe$ and $\rm CdTe$, we used the EPW package\cite{noffsinger2010epw} to explicitly calculate the dynamic electron-phonon scattering rates with the interpolation scheme based on the maximally localized Wannier functions (MLWFs). In the electron-phonon coupling calculations, we used the QUANTUM ESPRESSO package\cite{Giannozzi_2017} to evaluate the electronic band-structures and phonons in the Brillouin zone. A mesh grid of $\rm 20\times 20\times 20 $ was adopted for both materials with norm-conserving pseudopotentials. The kinetic energy cutoff for wave functions was set to 30 Ry. The kinetic energy cutoff for charge density and potential is set to 120 Ry. The total electron energy convergence threshold for self-consistency is $\rm 1\times 10^{-10}$ Ry. The crystal lattice is fully relaxed with a force threshold of $\rm 1 \times 10^{-4} eV/\AA$. Applying the density functional perturbation theory implemented in the QUANTUM ESPRESSO package\cite{Giannozzi_2017}, the phonon dispersion and the electron-phonon matrix elements are calculated on a coarse mesh of $\rm 6\times 6\times 6$. The electronic band structure, phonon dispersion relation, and electron-phonon scattering matrix elements are subsequently interpolated onto a fine mesh of $\rm 30\times 30\times 30$ using a MLWFs based scheme as implemented in the EPW package\cite{noffsinger2010epw}. We checked the convergence of the phonon scattering rates as a function of the fine sampling mesh density.

\begin{figure}
\includegraphics[width=0.7\linewidth,clip]{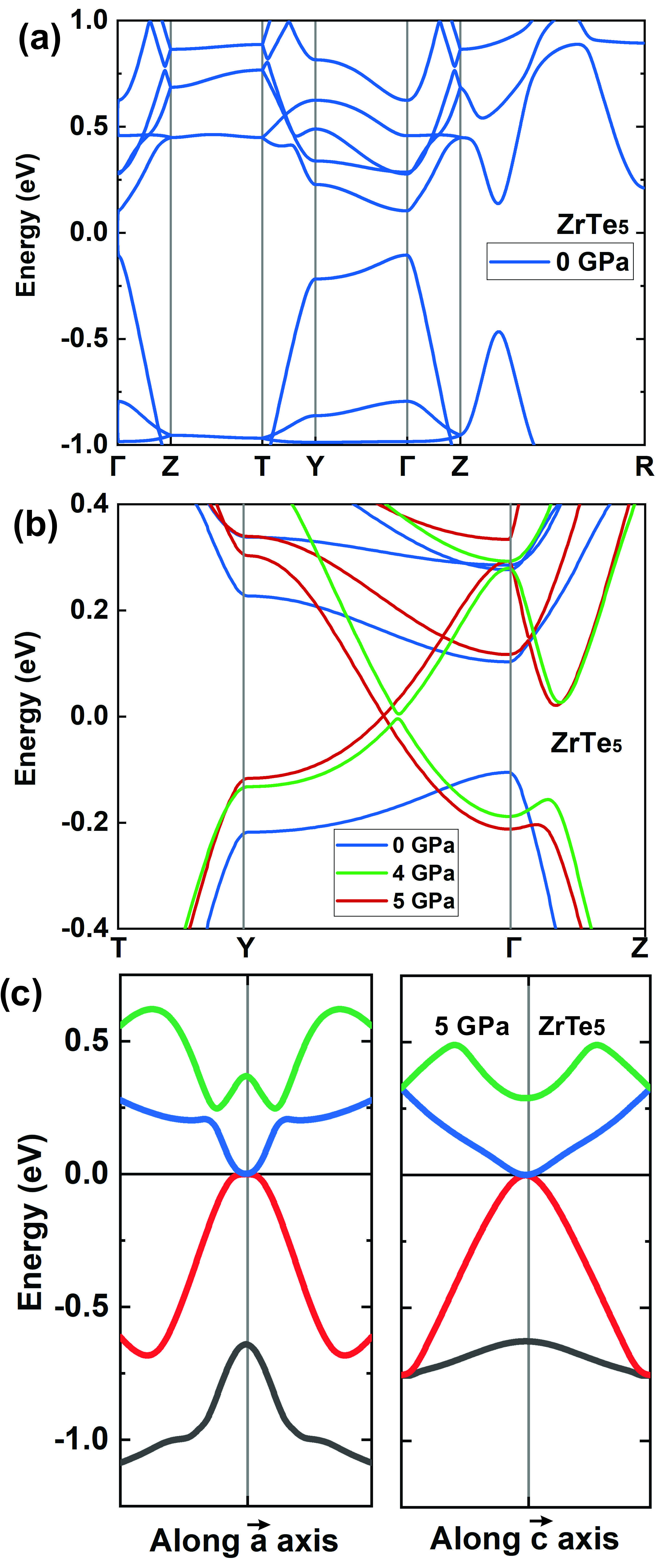}
\caption{\textbf{(a)} The calculated electronic band structure of $\rm ZrTe_5$ under zero pressure. \textbf{(b)} The calculated electronic band structure of $\rm ZrTe_5$ along the $\Gamma$-Y direction under different hydrostatic pressures. The band gap is closed when $\rm ZrTe_5$ is under 5 GPa pressure. \textbf{(c)} The quadratic electronic energy dispersion along the $a$ and $c$ axis near the band touching points in $\rm ZrTe_5$ under 5 GPa pressure.}
\label{fig:2}
\end{figure}

\section{Pressure Induced Topological Phase Transition in $\rm ZrTe_5$} 

$\rm ZrTe_5$ has been intensively studied recently due to its close proximity to a topological semimetal state. The nature of its ground state, however, remains controversial. While its 2D monolayer is predicted to be a quantum spin Hall insulator\cite{weng2014}, conflicting views about its 3D bulk band structure have been reported. Early studies based on transport\cite{liu2016zeeman,nair2018thermodynamic}, optical\cite{chen2015optical} and photoemission\cite{li2016chiral} measurements suggested that $\rm ZrTe_5$ is a topological Dirac semimetal with a single Dirac node at the Brillouin zone center. However, the vdW layered structure indicates that the Dirac node would lack the protection from additional crystalline symmetries as in other Dirac semimetals, e.g. $\rm Cd_3 As_2 $ and $\rm Na_3 Bi$. Other studies\cite{li2016experimental,chen2017spectroscopic,jiang2017landau,mutch2019evidence,xiong2017three,tang2019three,manzoni2016evidence,fan2017transition} concluded that $\rm ZrTe_5$ is a topological insulator with a small band gap at the zone center. In particular, Mutch and coworkers reported\cite{mutch2019evidence} that the bulk band gap in $\rm ZrTe_5$ can be closed by a small in-plane strain accompanying a TPT. The lack of a consensus so far implies the extreme sensitivity of the electronic bands in $\rm ZrTe_5$ to sample quality and external conditions.  

Our calculated electronic band structure of relaxed $\rm ZrTe_5$ is shown in Fig.~\ref{fig:2}(a), which is consistent with previous reports of a bulk band gap at the zone center $\Gamma$\cite{mutch2019evidence,manzoni2016evidence}. The band gap calculated in our work is roughly 150 meV and slightly higher than the experimental values below 100 meV. In addition to the in-plane strain induced TPT demonstrated by Mutch et al.\cite{mutch2019evidence}, we find that applying a hydrostatic pressure of 5 GPa can also drive a TPT and induce a semimetal phase (Fig.~\ref{fig:2}b). Our calculation is consistent with a previous report of pressure-induced superconductivity and metal-insulator transition in $\rm ZrTe_5$.\cite{zhou2016pressure} A recent study of pressure-dependent transport and infrared transmission study of $\rm ZrTe_5$ also signaled a pressure-induced band gap closing near 5 GPa\cite{santos2020probing}. Interestingly, in $\rm ZrTe_5$ under 5 GPa of hydrostatic pressure, the conduction band and the valence band touch at near half way along the $\mathrm{\Gamma-Y}$ direction (highlighted in Fig.~\ref{fig:1}a), which aligns with the layer-stacking $b$-axis in the conventional cell, agreeing with a previous calculation\cite{zhou2016pressure}. This is in contrast to the experimentally demonstrated strain-induced TPT\cite{mutch2019evidence} and the theoretically predicted lattice expansion induced TPT\cite{fan2017transition} in $\rm ZrTe_5$, where the band gap closes at the $\Gamma$ point. In addition, we find that the electronic energy dispersion near the touching points is linear along the out-of-plane direction ($\Gamma$-Y) with a Fermi velocity of approximately $3.3\times 10^5~ \rm{m/s}$. Along the in-plane directions, however, we find the energy dispersion near the touching points is anisotropic and quadratic (Fig.~\ref{fig:2}c). Along the $a$ ($c$) axis, the conduction band effective mass is 0.12 $m_0$ (0.5 $m_0$), and the valence band effective mass is 0.18 $m_0$ (0.28 $m_0$), respectively, where $m_0$ is the free electron mass. This characteristic differs from that of the low-energy bands near $\Gamma$ point under the ambient condition, where the in-plane dispersions are determined to be close to linear while the out-of-plane dispersion is quadratic\cite{martino2019two}. Our finding provides a possible explanation for the increased electrical resistivity\cite{santos2020probing} along the $a$-axis under hydrostatic pressure as the energy band contributing to the transport along $a$-axis evolves from being linear to being quadratic.

\begin{figure}
\includegraphics[width=0.7\linewidth,clip]{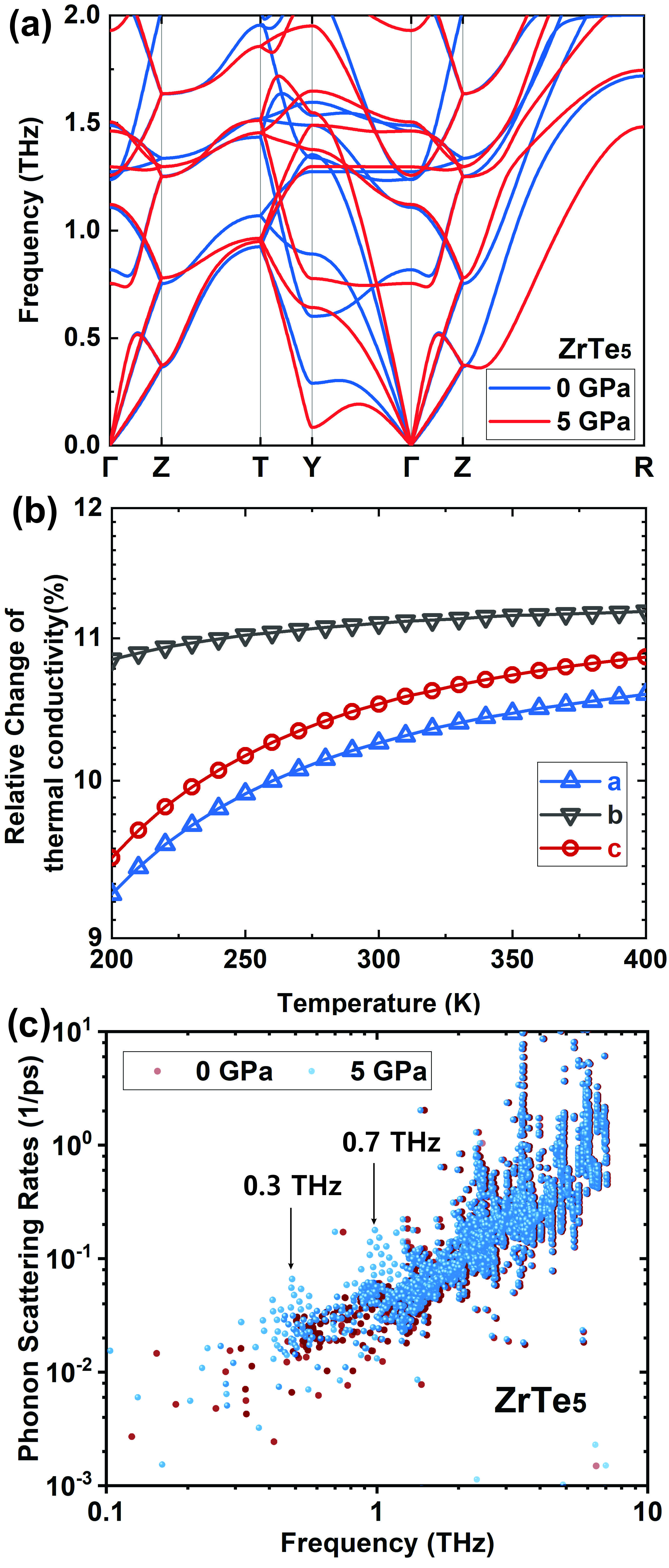}
\caption{\textbf{(a)} The calculated phonon dispersions of $\rm ZrTe_5$ under zero pressure and 5 GPa pressure. \textbf{(b)} The relative changes of the calculated lattice thermal conductivity of $\rm ZrTe_5$ along different directions under 5 GPa pressure compared to those under zero pressure. \textbf{(c)} The phonon-phonon scattering rates in $\rm ZrTe_5$ under zero pressure and 5 GPa pressure, highlighting the increased scattering near the softened phonons in pressured $\rm ZrTe_5$.}
\label{fig:3}
\end{figure}

The phonon dispersions of $\rm ZrTe_5$ under ambient condition and  5 GPa pressure are shown in Fig.~\ref{fig:3}a. The zero-pressure phonon dispersion is in good agreement with previous calculations\cite{zhu2018record,wang2018first}. A prominent feature is the significant softening of the transverse acoustic (TA) mode along the $\Gamma$-Y direction when $\rm ZrTe_5$ is subjected to the 5 GPa pressure, as well as a weaker kink and softening of the low-lying optical phonon at $\Gamma$ point. 

The observed phonon softening locations ($\Gamma$ and Y) suggest that they are instances of Kohn anomalies\cite{Kohn1959}. Kohn anomalies are the distortions of phonon dispersions observed in metals and semimetals caused by the resonance between the Fermi surface and certain phonon modes. Specifically, when two electronic states on the Fermi surface are parallelly connected (``nested'') by a phonon momentum $\mathbf{q}$, the polarizability $\Pi(\omega,\mathbf{q})$, which describes the collective response of the conduction electrons to an external disturbance with frequency $\omega$ and wavevector $\mathbf{q}$, becomes non-analytic. Since phononic vibrations are screened by the conduction electrons, the non-analyticity of the polarizability at certain phonon momentum $\mathbf{q}$ leads to abrupt changes of the phonon dispersion. The Fermi surface being two discrete nodes in $\rm ZrTe_5$ under 5 GPa pressure gives rise to two possible types of Kohn anomalies associated with intra-node or inter-node electron-phonon scatterings, which were similarly observed in graphene\cite{piscanec2004kohn,lazzeri2006nonadiabatic}. In the intra-node case, one electron within the vicinity of one Dirac node is scattered to another electronic state near the same node, mediated by a phonon with a wavevector $\bf{q} \thickapprox \bf{0}$. These processes are responsible for phonon anomalies near the Brillouin zone center, $\Gamma$. In the inter-node case, one electron close to one Dirac node is scattered to another electronic state near the other node, mediated by a phonon with a wavevector matching the distance between the two nodes, $\mathbf{q} \thickapprox 2\mathbf{k}_\mathrm{D}$, where $\mathbf{k}_\mathrm{D}$ marks the location of one Dirac node. In $\rm ZrTe_5$ under 5 GPa pressure, since the band touching point is located near the midpoint along the $\Gamma$-Y direction, the inter-node Kohn anomaly is expected to occur near Y, which is consistent with our calculation. 

\begin{figure*}%[h!]
\includegraphics[width=0.7\linewidth,clip]{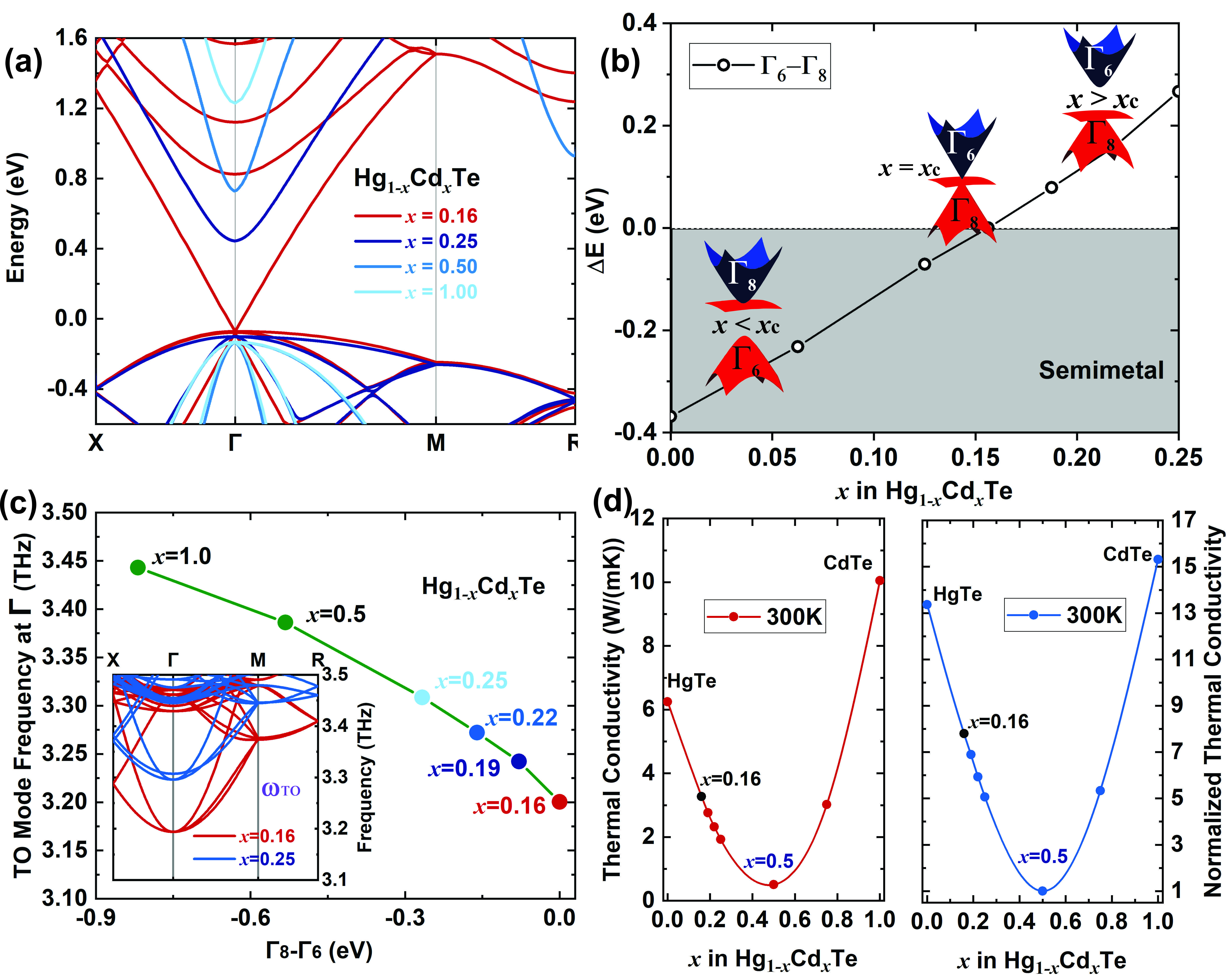}
\caption{\textbf{(a)} The calculated electronic band structure near $\Gamma$ of the $\rm Hg_{1-x}Cd_xTe$ alloys, showing the band closing at the critical composition $x=0.16$. \textbf{(b)} The evolution of the calculated band gap of $\rm Hg_{1-x}Cd_xTe$ as a function of the composition. \textbf{(c)} The evolution of the calculated TO phonon frequency in $\rm Hg_{1-x}Cd_xTe$ as a function of the band gap. The inset shows examples of the calculated optical phonon dispersion. \textbf{(d)} The calculated lattice thermal conductivity of $\rm Hg_{1-x}Cd_xTe$ is shown in the left panel. The calculated values are normalized by $v_s^2$ and shown in the right panel.} 
\label{fig:4}
\end{figure*}

Nguyen et al. showed analytically\cite{nguyen2020topological} that Kohn anomalies associated with three-dimensional Dirac nodes are caused by strong singularities in the electronic polarizability function $\Pi(\omega,\mathbf{q})$ that are similar to those in one-dimensional simple metals, where Kohn anomalies are known to give rise to structural instabilities (Peierls transitions)\cite{zhu2018record}. While the band touching points in $\rm ZrTe_5$ under 5 GPa pressure are not strictly Dirac nodes due to the quadratic dispersion along the $a$ and $c$ axis, the observed Kohn anomalies are strong and expected to significantly enlarge the available phase space for phonon scatterings and reduce the lattice thermal conductivity. The strong phonon softening at Y can also be responsible for the experimentally observed structural phase transition in $\rm ZrTe_5$ above 6 GPa\cite{zhou2016pressure}.

In Fig.~\ref{fig:3}(b), we show the relative change of the calculated lattice thermal conductivity of $\rm ZrTe_5$ induced by the 5 GPa hydrostatic pressure, as a function of the temperature. Under zero pressure, the lattice thermal conductivity along three axes are all different, with the one along the layer-stacking direction ($b$-axis) being the lowest (4 W/mK along $a$ axis, 0.4 W/mK along $b$ axis, and 1.8 W/mK along $c$ axis). Our results agree with the previous calculation and experiment by Zhu et al.\cite{zhu2018record}. Significant reductions of the lattice thermal conductivity above 10\% at room temperature along all three directions are observed when $\rm ZrTe_5$ is under 5 GPa pressure, particularly along the $b$ axis. More interestingly, we observed anomalous peaks in the phonon-phonon scattering rates of the low-frequency acoustic modes in pressured $\rm ZrTe_5$, as shown in Fig.~\ref{fig:3}c: one near 0.3 THz and the other near 0.7 THz, corresponding to the Kohn anomalies at $\rm Y$ and $\rm \Gamma$, respectively, providing strong evidence that the Kohn anomalies in pressured $\rm ZrTe_5$ contribute to its low lattice thermal conductivity. Our study of $\rm ZrTe_5$ suggests potential means to tune the thermal conductivity of topological materials through induced TPTs.

\section{Chemical composition induced topological phase transition in $\rm Hg_{1-x}Cd_xTe$}

\begin{figure}[ht]
\includegraphics[width=1.0\linewidth,clip]{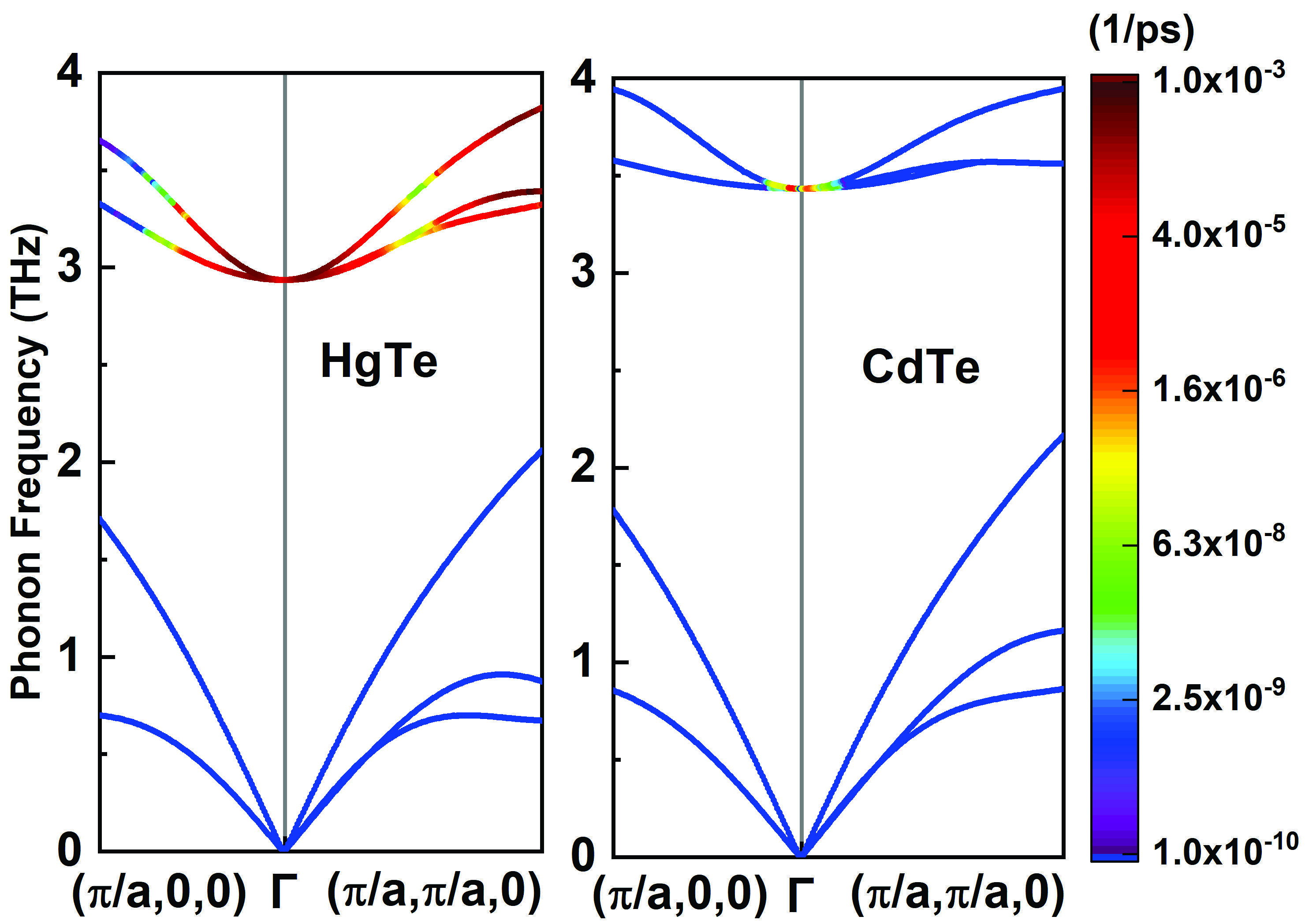}
\caption{The calculated mode-resolved phonon scattering rates due to electron-phonon interaction in HgTe (left) and CdTe (right). The color represents the scattering rates of the specific phonon modes.}
\label{fig:5}
\end{figure}

The HgTe/CdTe system is among the most studied model systems with nontrivial topological properties\cite{bernevig2006quantum}. HgTe is a semimetal with inverted electronics bands: the s-type $\Gamma_6$ band is below the p-type $\Gamma_8$ band in energy due to the strong spin-orbit coupling in Hg.\cite{zaheer2013spin} In comparison, CdTe is a finite-gap semiconductor with the normal band order ($\Gamma_6$ above $\Gamma_8$ in energy). Thus, by continuously substituting Cd for Hg, the electronic band structure of the $\rm Hg_{1-x}Cd_xTe$ alloy will evolve from an inverted semimetal to a normal insulator with distinct topological indices\cite{bernevig2006quantum}, going through a TPT where a three-dimensional Dirac dispersion near $\Gamma$ point emerges \cite{orlita2014observation,teppe2016temperature}. This transition is found experimentally to occur around $x=0.17$. The large tunability of the $\rm Hg_{1-x}Cd_xTe$ alloy has enabled its application in infrared photodetectors\cite{rogalski2005hgcdte}.

We show the calculated electronic band structure of $\rm Hg_{1-x}Cd_xTe$ alloys with different Cd concentration in Fig.~\ref{fig:4}a. The band gap of CdTe shrinks as the Cd concentration decreases, and completely closes when $x=0.16$ in our calculation. At the critical concentration when band closing occurs, linear conical bands emerge at the $\Gamma$ point. The evolution of $\rm Hg_{1-x}Cd_xTe$ between inverted semimetal to insulator phases is illustrated in Fig.~\ref{fig:4}b. The emergent band overlap at $\Gamma$ is expected to cause intranode Kohn anomalies that affect phonon modes near the $\Gamma$ point. As expected, our phonon calculation shows a clear drop of the transverse optical (TO) phonon frequency at $\Gamma$ as Cd concentration decreases. To eliminate the influence of the atomic mass difference between Hg and Cd on the TO phonon frequency, we plot the calculated TO phonon frequency normalized by a scaling factor $\frac{v_s(\rm HgTe)}{v_s(\rm Hg_{1-x}Cd_{x}Te)}$ in Fig.~\ref{fig:4}c, where $v_s$ is the speed of sound\cite{lee2014resonant}. Without the normalization, the TO phonon frequency would decrease even more with the decreasing Cd concentration. This decreasing trend qualitatively agrees with an infrared reflectivity measurement of $\rm Hg_{1-x}Cd_xTe$.\cite{sheregii2009temperature} A recent optical study of the $\rm Pb_{1-x}Sn_xSe$ system also showed the decreasing optical phonon frequency as the band gap shrinks to zero through a TPT\cite{wozny2020electron}. The TO phonon frequency further decreases past the TPT and into the semimetal phase, reaching a minimum in HgTe, indicating the persisting strong Kohn anomaly in the semimetal phase. 

We further evaluate the lattice thermal conductivity of $\rm Hg_{1-x}Cd_xTe$ at room temperature as a function of the Cd concentration, as shown in Fig.~\ref{fig:4}d. Our calculated result agrees well with experimental reports\cite{martyniuk2010mechanical}. The calculated lattice thermal conductivity is plotted in the left panel of Fig.~\ref{fig:4}d, while the lattice thermal conductivity normalized by $v_s^2$ is plotted in the right panel to  eliminate the impact of the Hg/Cd mass difference. Two major factors affect the lattice thermal conductivity of the $\rm Hg_{1-x}Cd_xTe$ alloys: strong phonon-defect scattering due to alloying and increased phonon-phonon scattering due to softened TO mode. The alloy effect typically leads to the minimum thermal conductivity near $x=0.5$, which is the case here in $\rm Hg_{1-x}Cd_xTe$. However, the normalized thermal conductivity of $\rm Hg_{1-x}Cd_xTe$ with small Cd concentration remains lower compared to compositions with small Hg concentration, such that the normalized thermal conductivity curve shown in Fig.~\ref{fig:4}d right panel is slightly skewed towards Hg-rich region. This result indicates the promise of $\rm Hg_{1-x}Cd_xTe$ near the TPT as a good thermoelectric material due to high mobility of the Dirac bands, strong electron-hole asymmetry for high Seebeck coefficients\cite{markov2018semi}, and low lattice thermal conductivity due to strong alloy scattering and soft optical phonons shown in this work.

To confirm that the origin of the softened TO phonons is Kohn anomalies induced by the strong electron-phonon interaction in the semimetal phase, we explicitly calculated the phonon scattering rates due to electron-phonon interaction\cite{liao2015significant,yue2019controlling} in HgTe and CdTe, as shown in Fig.~\ref{fig:5} (phonon dispersions are scaled by the speed of sound). It is clear that the TO phonon in HgTe is much more strongly scattered by electrons than that in CdTe. The TO phonon softening due to electron-phonon interaction in the $\rm Hg_{1-x}Cd_xTe$ system was previously discussed in terms of ``returnable'' electron-phonon interaction\cite{wozny2015influence}, meaning the impact of electron scatterings on phonons. Similar effects have been discussed in $\rm Pb_{1-x}Sn_xSe$ and $\rm Pb_{1-x}Sn_xTe$ systems\cite{kawamura1974dielectric,wozny2020electron} with small or zero band gaps, suggesting that the softened optical phonon is a universal feature of materials near TPTs. Intuitively, this is caused by more electron-phonon scattering channels when the electronic band gap is comparable to or smaller than the phonon energy scale.

\section{Conclusion}
In summary, we identified the existence of significant phonon softening near TPTs by examining two model systems: pressure induced TPT in $\rm ZrTe_5$ and chemical composition induced TPT in $\rm Hg_{1-x}Cd_xTe$ through first-principles simulations. We attributed the phonon softening to strong Kohn anomalies associated with the band gap closing at a TPT. We further evaluated the impact of the softened phonon modes on the thermal transport in these materials. Our study alludes to deep connections between the electronic structure and lattice dynamics and exemplifies the rich phonon physics in topological materials. Our results further suggest potential routes towards effective thermal conduction control and more efficient thermoelectric devices based on topological materials.    

\section{Acknowledgments}
This work is based on research supported by the U.S. Department of Energy, Office of Basic Energy Sciences, under the award number DE-SC0019244 and the UC Santa Barbara NSF Quantum Foundry funded via the Q-AMASE-i program under award DMR-1906325 (for studying topological materials), and the National Science Foundation under the award number CBET-1846927 (for studying phonon-electron scattering). Y.L. acknowledges the support from the Tsinghua Scholarship for Undergraduate Overseas Studies. This work used the Extreme Science and Engineering Discovery Environment (XSEDE), which is supported by National Science Foundation grant number ACI-1548562.

\renewcommand\refname{Reference}
\bibliography{refs}

%merlin.mbs apsrev4-1.bst 2010-07-25 4.21a (PWD, AO, DPC) hacked
%Control: key (0)
%Control: author (8) initials jnrlst
%Control: editor formatted (1) identically to author
%Control: production of article title (-1) disabled
%Control: page (0) single
%Control: year (1) truncated
%Control: production of eprint (0) enabled
\begin{thebibliography}{69}%
\makeatletter
\providecommand \@ifxundefined [1]{%
 \@ifx{#1\undefined}
}%
\providecommand \@ifnum [1]{%
 \ifnum #1\expandafter \@firstoftwo
 \else \expandafter \@secondoftwo
 \fi
}%
\providecommand \@ifx [1]{%
 \ifx #1\expandafter \@firstoftwo
 \else \expandafter \@secondoftwo
 \fi
}%
\providecommand \natexlab [1]{#1}%
\providecommand \enquote  [1]{``#1''}%
\providecommand \bibnamefont  [1]{#1}%
\providecommand \bibfnamefont [1]{#1}%
\providecommand \citenamefont [1]{#1}%
\providecommand \href@noop [0]{\@secondoftwo}%
\providecommand \href [0]{\begingroup \@sanitize@url \@href}%
\providecommand \@href[1]{\@@startlink{#1}\@@href}%
\providecommand \@@href[1]{\endgroup#1\@@endlink}%
\providecommand \@sanitize@url [0]{\catcode `\\12\catcode `\$12\catcode
  `\&12\catcode `\#12\catcode `\^12\catcode `\_12\catcode `\%12\relax}%
\providecommand \@@startlink[1]{}%
\providecommand \@@endlink[0]{}%
\providecommand \url  [0]{\begingroup\@sanitize@url \@url }%
\providecommand \@url [1]{\endgroup\@href {#1}{\urlprefix }}%
\providecommand \urlprefix  [0]{URL }%
\providecommand \Eprint [0]{\href }%
\providecommand \doibase [0]{http://dx.doi.org/}%
\providecommand \selectlanguage [0]{\@gobble}%
\providecommand \bibinfo  [0]{\@secondoftwo}%
\providecommand \bibfield  [0]{\@secondoftwo}%
\providecommand \translation [1]{[#1]}%
\providecommand \BibitemOpen [0]{}%
\providecommand \bibitemStop [0]{}%
\providecommand \bibitemNoStop [0]{.\EOS\space}%
\providecommand \EOS [0]{\spacefactor3000\relax}%
\providecommand \BibitemShut  [1]{\csname bibitem#1\endcsname}%
\let\auto@bib@innerbib\@empty
%</preamble>
\bibitem [{\citenamefont {Yan}\ and\ \citenamefont
  {Zhang}(2012)}]{yan2012topological}%
  \BibitemOpen
  \bibfield  {author} {\bibinfo {author} {\bibfnamefont {B.}~\bibnamefont
  {Yan}}\ and\ \bibinfo {author} {\bibfnamefont {S.-C.}\ \bibnamefont
  {Zhang}},\ }\href@noop {} {\bibfield  {journal} {\bibinfo  {journal} {Reports
  on Progress in Physics}\ }\textbf {\bibinfo {volume} {75}},\ \bibinfo {pages}
  {096501} (\bibinfo {year} {2012})}\BibitemShut {NoStop}%
\bibitem [{\citenamefont {Mutch}\ \emph {et~al.}(2019)\citenamefont {Mutch},
  \citenamefont {Chen}, \citenamefont {Went}, \citenamefont {Qian},
  \citenamefont {Wilson}, \citenamefont {Andreev}, \citenamefont {Chen},\ and\
  \citenamefont {Chu}}]{mutch2019evidence}%
  \BibitemOpen
  \bibfield  {author} {\bibinfo {author} {\bibfnamefont {J.}~\bibnamefont
  {Mutch}}, \bibinfo {author} {\bibfnamefont {W.-C.}\ \bibnamefont {Chen}},
  \bibinfo {author} {\bibfnamefont {P.}~\bibnamefont {Went}}, \bibinfo {author}
  {\bibfnamefont {T.}~\bibnamefont {Qian}}, \bibinfo {author} {\bibfnamefont
  {I.~Z.}\ \bibnamefont {Wilson}}, \bibinfo {author} {\bibfnamefont
  {A.}~\bibnamefont {Andreev}}, \bibinfo {author} {\bibfnamefont {C.-C.}\
  \bibnamefont {Chen}}, \ and\ \bibinfo {author} {\bibfnamefont {J.-H.}\
  \bibnamefont {Chu}},\ }\href@noop {} {\bibfield  {journal} {\bibinfo
  {journal} {Science Advances}\ }\textbf {\bibinfo {volume} {5}},\ \bibinfo
  {pages} {eaav9771} (\bibinfo {year} {2019})}\BibitemShut {NoStop}%
\bibitem [{\citenamefont {Orlita}\ \emph {et~al.}(2014)\citenamefont {Orlita},
  \citenamefont {Basko}, \citenamefont {Zholudev}, \citenamefont {Teppe},
  \citenamefont {Knap}, \citenamefont {Gavrilenko}, \citenamefont {Mikhailov},
  \citenamefont {Dvoretskii}, \citenamefont {Neugebauer}, \citenamefont
  {Faugeras} \emph {et~al.}}]{orlita2014observation}%
  \BibitemOpen
  \bibfield  {author} {\bibinfo {author} {\bibfnamefont {M.}~\bibnamefont
  {Orlita}}, \bibinfo {author} {\bibfnamefont {D.}~\bibnamefont {Basko}},
  \bibinfo {author} {\bibfnamefont {M.}~\bibnamefont {Zholudev}}, \bibinfo
  {author} {\bibfnamefont {F.}~\bibnamefont {Teppe}}, \bibinfo {author}
  {\bibfnamefont {W.}~\bibnamefont {Knap}}, \bibinfo {author} {\bibfnamefont
  {V.}~\bibnamefont {Gavrilenko}}, \bibinfo {author} {\bibfnamefont
  {N.}~\bibnamefont {Mikhailov}}, \bibinfo {author} {\bibfnamefont
  {S.}~\bibnamefont {Dvoretskii}}, \bibinfo {author} {\bibfnamefont
  {P.}~\bibnamefont {Neugebauer}}, \bibinfo {author} {\bibfnamefont
  {C.}~\bibnamefont {Faugeras}},  \emph {et~al.},\ }\href@noop {} {\bibfield
  {journal} {\bibinfo  {journal} {Nature Physics}\ }\textbf {\bibinfo {volume}
  {10}},\ \bibinfo {pages} {233} (\bibinfo {year} {2014})}\BibitemShut
  {NoStop}%
\bibitem [{\citenamefont {Teppe}\ \emph {et~al.}(2016)\citenamefont {Teppe},
  \citenamefont {Marcinkiewicz}, \citenamefont {Krishtopenko}, \citenamefont
  {Ruffenach}, \citenamefont {Consejo}, \citenamefont {Kadykov}, \citenamefont
  {Desrat}, \citenamefont {But}, \citenamefont {Knap}, \citenamefont {Ludwig}
  \emph {et~al.}}]{teppe2016temperature}%
  \BibitemOpen
  \bibfield  {author} {\bibinfo {author} {\bibfnamefont {F.}~\bibnamefont
  {Teppe}}, \bibinfo {author} {\bibfnamefont {M.}~\bibnamefont
  {Marcinkiewicz}}, \bibinfo {author} {\bibfnamefont {S.}~\bibnamefont
  {Krishtopenko}}, \bibinfo {author} {\bibfnamefont {S.}~\bibnamefont
  {Ruffenach}}, \bibinfo {author} {\bibfnamefont {C.}~\bibnamefont {Consejo}},
  \bibinfo {author} {\bibfnamefont {A.}~\bibnamefont {Kadykov}}, \bibinfo
  {author} {\bibfnamefont {W.}~\bibnamefont {Desrat}}, \bibinfo {author}
  {\bibfnamefont {D.}~\bibnamefont {But}}, \bibinfo {author} {\bibfnamefont
  {W.}~\bibnamefont {Knap}}, \bibinfo {author} {\bibfnamefont {J.}~\bibnamefont
  {Ludwig}},  \emph {et~al.},\ }\href@noop {} {\bibfield  {journal} {\bibinfo
  {journal} {Nature Communications}\ }\textbf {\bibinfo {volume} {7}},\
  \bibinfo {pages} {1} (\bibinfo {year} {2016})}\BibitemShut {NoStop}%
\bibitem [{\citenamefont {Ezawa}(2013)}]{ezawa2013photoinduced}%
  \BibitemOpen
  \bibfield  {author} {\bibinfo {author} {\bibfnamefont {M.}~\bibnamefont
  {Ezawa}},\ }\href@noop {} {\bibfield  {journal} {\bibinfo  {journal}
  {Physical Review Letters}\ }\textbf {\bibinfo {volume} {110}},\ \bibinfo
  {pages} {026603} (\bibinfo {year} {2013})}\BibitemShut {NoStop}%
\bibitem [{\citenamefont {Collins}\ \emph {et~al.}(2018)\citenamefont
  {Collins}, \citenamefont {Tadich}, \citenamefont {Wu}, \citenamefont {Gomes},
  \citenamefont {Rodrigues}, \citenamefont {Liu}, \citenamefont {Hellerstedt},
  \citenamefont {Ryu}, \citenamefont {Tang}, \citenamefont {Mo} \emph
  {et~al.}}]{collins2018electric}%
  \BibitemOpen
  \bibfield  {author} {\bibinfo {author} {\bibfnamefont {J.~L.}\ \bibnamefont
  {Collins}}, \bibinfo {author} {\bibfnamefont {A.}~\bibnamefont {Tadich}},
  \bibinfo {author} {\bibfnamefont {W.}~\bibnamefont {Wu}}, \bibinfo {author}
  {\bibfnamefont {L.~C.}\ \bibnamefont {Gomes}}, \bibinfo {author}
  {\bibfnamefont {J.~N.}\ \bibnamefont {Rodrigues}}, \bibinfo {author}
  {\bibfnamefont {C.}~\bibnamefont {Liu}}, \bibinfo {author} {\bibfnamefont
  {J.}~\bibnamefont {Hellerstedt}}, \bibinfo {author} {\bibfnamefont
  {H.}~\bibnamefont {Ryu}}, \bibinfo {author} {\bibfnamefont {S.}~\bibnamefont
  {Tang}}, \bibinfo {author} {\bibfnamefont {S.-K.}\ \bibnamefont {Mo}},  \emph
  {et~al.},\ }\href@noop {} {\bibfield  {journal} {\bibinfo  {journal}
  {Nature}\ }\textbf {\bibinfo {volume} {564}},\ \bibinfo {pages} {390}
  (\bibinfo {year} {2018})}\BibitemShut {NoStop}%
\bibitem [{\citenamefont {Zhu}\ \emph {et~al.}(2012)\citenamefont {Zhu},
  \citenamefont {Cheng},\ and\ \citenamefont
  {Schwingenschl{\"o}gl}}]{zhu2012topological}%
  \BibitemOpen
  \bibfield  {author} {\bibinfo {author} {\bibfnamefont {Z.}~\bibnamefont
  {Zhu}}, \bibinfo {author} {\bibfnamefont {Y.}~\bibnamefont {Cheng}}, \ and\
  \bibinfo {author} {\bibfnamefont {U.}~\bibnamefont {Schwingenschl{\"o}gl}},\
  }\href@noop {} {\bibfield  {journal} {\bibinfo  {journal} {Physical Review
  Letters}\ }\textbf {\bibinfo {volume} {108}},\ \bibinfo {pages} {266805}
  (\bibinfo {year} {2012})}\BibitemShut {NoStop}%
\bibitem [{\citenamefont {Fan}\ \emph {et~al.}(2017{\natexlab{a}})\citenamefont
  {Fan}, \citenamefont {Liang}, \citenamefont {Chen}, \citenamefont {Yao},\
  and\ \citenamefont {Zhou}}]{fan2017transition}%
  \BibitemOpen
  \bibfield  {author} {\bibinfo {author} {\bibfnamefont {Z.}~\bibnamefont
  {Fan}}, \bibinfo {author} {\bibfnamefont {Q.-F.}\ \bibnamefont {Liang}},
  \bibinfo {author} {\bibfnamefont {Y.}~\bibnamefont {Chen}}, \bibinfo {author}
  {\bibfnamefont {S.-H.}\ \bibnamefont {Yao}}, \ and\ \bibinfo {author}
  {\bibfnamefont {J.}~\bibnamefont {Zhou}},\ }\href@noop {} {\bibfield
  {journal} {\bibinfo  {journal} {Scientific Reports}\ }\textbf {\bibinfo
  {volume} {7}},\ \bibinfo {pages} {45667} (\bibinfo {year}
  {2017}{\natexlab{a}})}\BibitemShut {NoStop}%
\bibitem [{\citenamefont {Bernevig}\ \emph {et~al.}(2006)\citenamefont
  {Bernevig}, \citenamefont {Hughes},\ and\ \citenamefont
  {Zhang}}]{bernevig2006quantum}%
  \BibitemOpen
  \bibfield  {author} {\bibinfo {author} {\bibfnamefont {B.~A.}\ \bibnamefont
  {Bernevig}}, \bibinfo {author} {\bibfnamefont {T.~L.}\ \bibnamefont
  {Hughes}}, \ and\ \bibinfo {author} {\bibfnamefont {S.-C.}\ \bibnamefont
  {Zhang}},\ }\href@noop {} {\bibfield  {journal} {\bibinfo  {journal}
  {Science}\ }\textbf {\bibinfo {volume} {314}},\ \bibinfo {pages} {1757}
  (\bibinfo {year} {2006})}\BibitemShut {NoStop}%
\bibitem [{\citenamefont {Xu}\ \emph {et~al.}(2011)\citenamefont {Xu},
  \citenamefont {Xia}, \citenamefont {Wray}, \citenamefont {Jia}, \citenamefont
  {Meier}, \citenamefont {Dil}, \citenamefont {Osterwalder}, \citenamefont
  {Slomski}, \citenamefont {Bansil}, \citenamefont {Lin} \emph
  {et~al.}}]{xu2011topological}%
  \BibitemOpen
  \bibfield  {author} {\bibinfo {author} {\bibfnamefont {S.-Y.}\ \bibnamefont
  {Xu}}, \bibinfo {author} {\bibfnamefont {Y.}~\bibnamefont {Xia}}, \bibinfo
  {author} {\bibfnamefont {L.}~\bibnamefont {Wray}}, \bibinfo {author}
  {\bibfnamefont {S.}~\bibnamefont {Jia}}, \bibinfo {author} {\bibfnamefont
  {F.}~\bibnamefont {Meier}}, \bibinfo {author} {\bibfnamefont
  {J.}~\bibnamefont {Dil}}, \bibinfo {author} {\bibfnamefont {J.}~\bibnamefont
  {Osterwalder}}, \bibinfo {author} {\bibfnamefont {B.}~\bibnamefont
  {Slomski}}, \bibinfo {author} {\bibfnamefont {A.}~\bibnamefont {Bansil}},
  \bibinfo {author} {\bibfnamefont {H.}~\bibnamefont {Lin}},  \emph {et~al.},\
  }\href@noop {} {\bibfield  {journal} {\bibinfo  {journal} {Science}\ }\textbf
  {\bibinfo {volume} {332}},\ \bibinfo {pages} {560} (\bibinfo {year}
  {2011})}\BibitemShut {NoStop}%
\bibitem [{\citenamefont {Wu}\ \emph {et~al.}(2013)\citenamefont {Wu},
  \citenamefont {Brahlek}, \citenamefont {Aguilar}, \citenamefont {Stier},
  \citenamefont {Morris}, \citenamefont {Lubashevsky}, \citenamefont {Bilbro},
  \citenamefont {Bansal}, \citenamefont {Oh},\ and\ \citenamefont
  {Armitage}}]{wu2013sudden}%
  \BibitemOpen
  \bibfield  {author} {\bibinfo {author} {\bibfnamefont {L.}~\bibnamefont
  {Wu}}, \bibinfo {author} {\bibfnamefont {M.}~\bibnamefont {Brahlek}},
  \bibinfo {author} {\bibfnamefont {R.~V.}\ \bibnamefont {Aguilar}}, \bibinfo
  {author} {\bibfnamefont {A.}~\bibnamefont {Stier}}, \bibinfo {author}
  {\bibfnamefont {C.}~\bibnamefont {Morris}}, \bibinfo {author} {\bibfnamefont
  {Y.}~\bibnamefont {Lubashevsky}}, \bibinfo {author} {\bibfnamefont
  {L.}~\bibnamefont {Bilbro}}, \bibinfo {author} {\bibfnamefont
  {N.}~\bibnamefont {Bansal}}, \bibinfo {author} {\bibfnamefont
  {S.}~\bibnamefont {Oh}}, \ and\ \bibinfo {author} {\bibfnamefont
  {N.}~\bibnamefont {Armitage}},\ }\href@noop {} {\bibfield  {journal}
  {\bibinfo  {journal} {Nature Physics}\ }\textbf {\bibinfo {volume} {9}},\
  \bibinfo {pages} {410} (\bibinfo {year} {2013})}\BibitemShut {NoStop}%
\bibitem [{\citenamefont {Dziawa}\ \emph {et~al.}(2012)\citenamefont {Dziawa},
  \citenamefont {Kowalski}, \citenamefont {Dybko}, \citenamefont {Buczko},
  \citenamefont {Szczerbakow}, \citenamefont {Szot}, \citenamefont
  {{\L}usakowska}, \citenamefont {Balasubramanian}, \citenamefont {Wojek},
  \citenamefont {Berntsen} \emph {et~al.}}]{dziawa2012topological}%
  \BibitemOpen
  \bibfield  {author} {\bibinfo {author} {\bibfnamefont {P.}~\bibnamefont
  {Dziawa}}, \bibinfo {author} {\bibfnamefont {B.}~\bibnamefont {Kowalski}},
  \bibinfo {author} {\bibfnamefont {K.}~\bibnamefont {Dybko}}, \bibinfo
  {author} {\bibfnamefont {R.}~\bibnamefont {Buczko}}, \bibinfo {author}
  {\bibfnamefont {A.}~\bibnamefont {Szczerbakow}}, \bibinfo {author}
  {\bibfnamefont {M.}~\bibnamefont {Szot}}, \bibinfo {author} {\bibfnamefont
  {E.}~\bibnamefont {{\L}usakowska}}, \bibinfo {author} {\bibfnamefont
  {T.}~\bibnamefont {Balasubramanian}}, \bibinfo {author} {\bibfnamefont
  {B.~M.}\ \bibnamefont {Wojek}}, \bibinfo {author} {\bibfnamefont
  {M.}~\bibnamefont {Berntsen}},  \emph {et~al.},\ }\href@noop {} {\bibfield
  {journal} {\bibinfo  {journal} {Nature Materials}\ }\textbf {\bibinfo
  {volume} {11}},\ \bibinfo {pages} {1023} (\bibinfo {year}
  {2012})}\BibitemShut {NoStop}%
\bibitem [{\citenamefont {Xu}\ \emph {et~al.}(2012)\citenamefont {Xu},
  \citenamefont {Liu}, \citenamefont {Alidoust}, \citenamefont {Neupane},
  \citenamefont {Qian}, \citenamefont {Belopolski}, \citenamefont {Denlinger},
  \citenamefont {Wang}, \citenamefont {Lin}, \citenamefont {Wray} \emph
  {et~al.}}]{xu2012observation}%
  \BibitemOpen
  \bibfield  {author} {\bibinfo {author} {\bibfnamefont {S.-Y.}\ \bibnamefont
  {Xu}}, \bibinfo {author} {\bibfnamefont {C.}~\bibnamefont {Liu}}, \bibinfo
  {author} {\bibfnamefont {N.}~\bibnamefont {Alidoust}}, \bibinfo {author}
  {\bibfnamefont {M.}~\bibnamefont {Neupane}}, \bibinfo {author} {\bibfnamefont
  {D.}~\bibnamefont {Qian}}, \bibinfo {author} {\bibfnamefont {I.}~\bibnamefont
  {Belopolski}}, \bibinfo {author} {\bibfnamefont {J.}~\bibnamefont
  {Denlinger}}, \bibinfo {author} {\bibfnamefont {Y.}~\bibnamefont {Wang}},
  \bibinfo {author} {\bibfnamefont {H.}~\bibnamefont {Lin}}, \bibinfo {author}
  {\bibfnamefont {L.~a.}\ \bibnamefont {Wray}},  \emph {et~al.},\ }\href@noop
  {} {\bibfield  {journal} {\bibinfo  {journal} {Nature Communications}\
  }\textbf {\bibinfo {volume} {3}},\ \bibinfo {pages} {1} (\bibinfo {year}
  {2012})}\BibitemShut {NoStop}%
\bibitem [{\citenamefont {Xu}\ \emph {et~al.}(2018)\citenamefont {Xu},
  \citenamefont {Zhao}, \citenamefont {Marsik}, \citenamefont {Sheveleva},
  \citenamefont {Lyzwa}, \citenamefont {Dai}, \citenamefont {Chen},
  \citenamefont {Qiu},\ and\ \citenamefont {Bernhard}}]{xu2018temperature}%
  \BibitemOpen
  \bibfield  {author} {\bibinfo {author} {\bibfnamefont {B.}~\bibnamefont
  {Xu}}, \bibinfo {author} {\bibfnamefont {L.}~\bibnamefont {Zhao}}, \bibinfo
  {author} {\bibfnamefont {P.}~\bibnamefont {Marsik}}, \bibinfo {author}
  {\bibfnamefont {E.}~\bibnamefont {Sheveleva}}, \bibinfo {author}
  {\bibfnamefont {F.}~\bibnamefont {Lyzwa}}, \bibinfo {author} {\bibfnamefont
  {Y.}~\bibnamefont {Dai}}, \bibinfo {author} {\bibfnamefont {G.}~\bibnamefont
  {Chen}}, \bibinfo {author} {\bibfnamefont {X.}~\bibnamefont {Qiu}}, \ and\
  \bibinfo {author} {\bibfnamefont {C.}~\bibnamefont {Bernhard}},\ }\href@noop
  {} {\bibfield  {journal} {\bibinfo  {journal} {Physical Review Letters}\
  }\textbf {\bibinfo {volume} {121}},\ \bibinfo {pages} {187401} (\bibinfo
  {year} {2018})}\BibitemShut {NoStop}%
\bibitem [{\citenamefont {Chen}\ \emph {et~al.}(2019)\citenamefont {Chen},
  \citenamefont {Chen}, \citenamefont {Li}, \citenamefont {Zhang},
  \citenamefont {Struzhkin}, \citenamefont {Goncharov}, \citenamefont {Ren},\
  and\ \citenamefont {Chen}}]{chen2019enhancement}%
  \BibitemOpen
  \bibfield  {author} {\bibinfo {author} {\bibfnamefont {L.-C.}\ \bibnamefont
  {Chen}}, \bibinfo {author} {\bibfnamefont {P.-Q.}\ \bibnamefont {Chen}},
  \bibinfo {author} {\bibfnamefont {W.-J.}\ \bibnamefont {Li}}, \bibinfo
  {author} {\bibfnamefont {Q.}~\bibnamefont {Zhang}}, \bibinfo {author}
  {\bibfnamefont {V.~V.}\ \bibnamefont {Struzhkin}}, \bibinfo {author}
  {\bibfnamefont {A.~F.}\ \bibnamefont {Goncharov}}, \bibinfo {author}
  {\bibfnamefont {Z.}~\bibnamefont {Ren}}, \ and\ \bibinfo {author}
  {\bibfnamefont {X.-J.}\ \bibnamefont {Chen}},\ }\href@noop {} {\bibfield
  {journal} {\bibinfo  {journal} {Nature Materials}\ }\textbf {\bibinfo
  {volume} {18}},\ \bibinfo {pages} {1321} (\bibinfo {year}
  {2019})}\BibitemShut {NoStop}%
\bibitem [{\citenamefont {Wang}\ \emph {et~al.}(2013)\citenamefont {Wang},
  \citenamefont {Weng}, \citenamefont {Wu}, \citenamefont {Dai},\ and\
  \citenamefont {Fang}}]{wang2013-Cd3As2}%
  \BibitemOpen
  \bibfield  {author} {\bibinfo {author} {\bibfnamefont {Z.}~\bibnamefont
  {Wang}}, \bibinfo {author} {\bibfnamefont {H.}~\bibnamefont {Weng}}, \bibinfo
  {author} {\bibfnamefont {Q.}~\bibnamefont {Wu}}, \bibinfo {author}
  {\bibfnamefont {X.}~\bibnamefont {Dai}}, \ and\ \bibinfo {author}
  {\bibfnamefont {Z.}~\bibnamefont {Fang}},\ }\href@noop {} {\bibfield
  {journal} {\bibinfo  {journal} {Physical Review B}\ }\textbf {\bibinfo
  {volume} {88}},\ \bibinfo {pages} {125427} (\bibinfo {year}
  {2013})}\BibitemShut {NoStop}%
\bibitem [{\citenamefont {Ominato}\ and\ \citenamefont
  {Koshino}(2014)}]{Ominato2014}%
  \BibitemOpen
  \bibfield  {author} {\bibinfo {author} {\bibfnamefont {Y.}~\bibnamefont
  {Ominato}}\ and\ \bibinfo {author} {\bibfnamefont {M.}~\bibnamefont
  {Koshino}},\ }\href@noop {} {\bibfield  {journal} {\bibinfo  {journal}
  {Physical Review B}\ }\textbf {\bibinfo {volume} {89}},\ \bibinfo {pages}
  {054202} (\bibinfo {year} {2014})}\BibitemShut {NoStop}%
\bibitem [{\citenamefont {Neupane}\ \emph {et~al.}(2014)\citenamefont
  {Neupane}, \citenamefont {Xu}, \citenamefont {Sankar}, \citenamefont
  {Alidoust}, \citenamefont {Bian}, \citenamefont {Liu}, \citenamefont
  {Belopolski}, \citenamefont {Chang}, \citenamefont {Jeng}, \citenamefont
  {Lin} \emph {et~al.}}]{neupane2014observation}%
  \BibitemOpen
  \bibfield  {author} {\bibinfo {author} {\bibfnamefont {M.}~\bibnamefont
  {Neupane}}, \bibinfo {author} {\bibfnamefont {S.-Y.}\ \bibnamefont {Xu}},
  \bibinfo {author} {\bibfnamefont {R.}~\bibnamefont {Sankar}}, \bibinfo
  {author} {\bibfnamefont {N.}~\bibnamefont {Alidoust}}, \bibinfo {author}
  {\bibfnamefont {G.}~\bibnamefont {Bian}}, \bibinfo {author} {\bibfnamefont
  {C.}~\bibnamefont {Liu}}, \bibinfo {author} {\bibfnamefont {I.}~\bibnamefont
  {Belopolski}}, \bibinfo {author} {\bibfnamefont {T.-R.}\ \bibnamefont
  {Chang}}, \bibinfo {author} {\bibfnamefont {H.-T.}\ \bibnamefont {Jeng}},
  \bibinfo {author} {\bibfnamefont {H.}~\bibnamefont {Lin}},  \emph {et~al.},\
  }\href@noop {} {\bibfield  {journal} {\bibinfo  {journal} {Nature
  Communications}\ }\textbf {\bibinfo {volume} {5}},\ \bibinfo {pages} {3786}
  (\bibinfo {year} {2014})}\BibitemShut {NoStop}%
\bibitem [{\citenamefont {Schumann}\ \emph {et~al.}(2016)\citenamefont
  {Schumann}, \citenamefont {Goyal}, \citenamefont {Kim},\ and\ \citenamefont
  {Stemmer}}]{schumann2016molecular}%
  \BibitemOpen
  \bibfield  {author} {\bibinfo {author} {\bibfnamefont {T.}~\bibnamefont
  {Schumann}}, \bibinfo {author} {\bibfnamefont {M.}~\bibnamefont {Goyal}},
  \bibinfo {author} {\bibfnamefont {H.}~\bibnamefont {Kim}}, \ and\ \bibinfo
  {author} {\bibfnamefont {S.}~\bibnamefont {Stemmer}},\ }\href@noop {}
  {\bibfield  {journal} {\bibinfo  {journal} {APL Materials}\ }\textbf
  {\bibinfo {volume} {4}},\ \bibinfo {pages} {126110} (\bibinfo {year}
  {2016})}\BibitemShut {NoStop}%
\bibitem [{\citenamefont {Wan}\ \emph {et~al.}(2011)\citenamefont {Wan},
  \citenamefont {Turner}, \citenamefont {Vishwanath},\ and\ \citenamefont
  {Savrasov}}]{Wan2011-Fermi-arc}%
  \BibitemOpen
  \bibfield  {author} {\bibinfo {author} {\bibfnamefont {X.}~\bibnamefont
  {Wan}}, \bibinfo {author} {\bibfnamefont {A.~M.}\ \bibnamefont {Turner}},
  \bibinfo {author} {\bibfnamefont {A.}~\bibnamefont {Vishwanath}}, \ and\
  \bibinfo {author} {\bibfnamefont {S.~Y.}\ \bibnamefont {Savrasov}},\
  }\href@noop {} {\bibfield  {journal} {\bibinfo  {journal} {Physical Review
  B}\ }\textbf {\bibinfo {volume} {83}},\ \bibinfo {pages} {205101} (\bibinfo
  {year} {2011})}\BibitemShut {NoStop}%
\bibitem [{\citenamefont {Xu}\ \emph {et~al.}(2015)\citenamefont {Xu},
  \citenamefont {Liu}, \citenamefont {Kushwaha}, \citenamefont {Sankar},
  \citenamefont {Krizan}, \citenamefont {Belopolski}, \citenamefont {Neupane},
  \citenamefont {Bian}, \citenamefont {Alidoust}, \citenamefont {Chang},
  \citenamefont {Jeng}, \citenamefont {Huang}, \citenamefont {Tsai},
  \citenamefont {Lin}, \citenamefont {Shibayev}, \citenamefont {Chou},
  \citenamefont {Cava},\ and\ \citenamefont {Hasan}}]{Xu2015-Fermi-arc}%
  \BibitemOpen
  \bibfield  {author} {\bibinfo {author} {\bibfnamefont {S.-Y.}\ \bibnamefont
  {Xu}}, \bibinfo {author} {\bibfnamefont {C.}~\bibnamefont {Liu}}, \bibinfo
  {author} {\bibfnamefont {S.~K.}\ \bibnamefont {Kushwaha}}, \bibinfo {author}
  {\bibfnamefont {R.}~\bibnamefont {Sankar}}, \bibinfo {author} {\bibfnamefont
  {J.~W.}\ \bibnamefont {Krizan}}, \bibinfo {author} {\bibfnamefont
  {I.}~\bibnamefont {Belopolski}}, \bibinfo {author} {\bibfnamefont
  {M.}~\bibnamefont {Neupane}}, \bibinfo {author} {\bibfnamefont
  {G.}~\bibnamefont {Bian}}, \bibinfo {author} {\bibfnamefont {N.}~\bibnamefont
  {Alidoust}}, \bibinfo {author} {\bibfnamefont {T.-R.}\ \bibnamefont {Chang}},
  \bibinfo {author} {\bibfnamefont {H.-T.}\ \bibnamefont {Jeng}}, \bibinfo
  {author} {\bibfnamefont {C.-Y.}\ \bibnamefont {Huang}}, \bibinfo {author}
  {\bibfnamefont {W.-F.}\ \bibnamefont {Tsai}}, \bibinfo {author}
  {\bibfnamefont {H.}~\bibnamefont {Lin}}, \bibinfo {author} {\bibfnamefont
  {P.~P.}\ \bibnamefont {Shibayev}}, \bibinfo {author} {\bibfnamefont {F.-C.}\
  \bibnamefont {Chou}}, \bibinfo {author} {\bibfnamefont {R.~J.}\ \bibnamefont
  {Cava}}, \ and\ \bibinfo {author} {\bibfnamefont {M.~Z.}\ \bibnamefont
  {Hasan}},\ }\href@noop {} {\bibfield  {journal} {\bibinfo  {journal}
  {Science}\ }\textbf {\bibinfo {volume} {347}},\ \bibinfo {pages} {294}
  (\bibinfo {year} {2015})}\BibitemShut {NoStop}%
\bibitem [{\citenamefont {Moll}\ \emph {et~al.}(2016)\citenamefont {Moll},
  \citenamefont {Nair}, \citenamefont {Helm}, \citenamefont {Potter},
  \citenamefont {Kimchi}, \citenamefont {Vishwanath},\ and\ \citenamefont
  {Analytis}}]{moll2016transport}%
  \BibitemOpen
  \bibfield  {author} {\bibinfo {author} {\bibfnamefont {P.~J.}\ \bibnamefont
  {Moll}}, \bibinfo {author} {\bibfnamefont {N.~L.}\ \bibnamefont {Nair}},
  \bibinfo {author} {\bibfnamefont {T.}~\bibnamefont {Helm}}, \bibinfo {author}
  {\bibfnamefont {A.~C.}\ \bibnamefont {Potter}}, \bibinfo {author}
  {\bibfnamefont {I.}~\bibnamefont {Kimchi}}, \bibinfo {author} {\bibfnamefont
  {A.}~\bibnamefont {Vishwanath}}, \ and\ \bibinfo {author} {\bibfnamefont
  {J.~G.}\ \bibnamefont {Analytis}},\ }\href@noop {} {\bibfield  {journal}
  {\bibinfo  {journal} {Nature}\ }\textbf {\bibinfo {volume} {535}},\ \bibinfo
  {pages} {266} (\bibinfo {year} {2016})}\BibitemShut {NoStop}%
\bibitem [{\citenamefont {Chorsi}\ \emph {et~al.}(2019)\citenamefont {Chorsi},
  \citenamefont {Yue}, \citenamefont {Iyer}, \citenamefont {Goyal},
  \citenamefont {Schumann}, \citenamefont {Stemmer}, \citenamefont {Liao},\
  and\ \citenamefont {Schuller}}]{chorsi2019widely}%
  \BibitemOpen
  \bibfield  {author} {\bibinfo {author} {\bibfnamefont {H.~T.}\ \bibnamefont
  {Chorsi}}, \bibinfo {author} {\bibfnamefont {S.}~\bibnamefont {Yue}},
  \bibinfo {author} {\bibfnamefont {P.~P.}\ \bibnamefont {Iyer}}, \bibinfo
  {author} {\bibfnamefont {M.}~\bibnamefont {Goyal}}, \bibinfo {author}
  {\bibfnamefont {T.}~\bibnamefont {Schumann}}, \bibinfo {author}
  {\bibfnamefont {S.}~\bibnamefont {Stemmer}}, \bibinfo {author} {\bibfnamefont
  {B.}~\bibnamefont {Liao}}, \ and\ \bibinfo {author} {\bibfnamefont {J.~A.}\
  \bibnamefont {Schuller}},\ }\href@noop {} {\bibfield  {journal} {\bibinfo
  {journal} {arXiv preprint arXiv:1907.12105}\ } (\bibinfo {year}
  {2019})}\BibitemShut {NoStop}%
\bibitem [{\citenamefont {Yue}\ \emph {et~al.}(2019{\natexlab{a}})\citenamefont
  {Yue}, \citenamefont {Chorsi}, \citenamefont {Goyal}, \citenamefont
  {Schumann}, \citenamefont {Yang}, \citenamefont {Xu}, \citenamefont {Deng},
  \citenamefont {Stemmer}, \citenamefont {Schuller},\ and\ \citenamefont
  {Liao}}]{yue2019soft}%
  \BibitemOpen
  \bibfield  {author} {\bibinfo {author} {\bibfnamefont {S.}~\bibnamefont
  {Yue}}, \bibinfo {author} {\bibfnamefont {H.~T.}\ \bibnamefont {Chorsi}},
  \bibinfo {author} {\bibfnamefont {M.}~\bibnamefont {Goyal}}, \bibinfo
  {author} {\bibfnamefont {T.}~\bibnamefont {Schumann}}, \bibinfo {author}
  {\bibfnamefont {R.}~\bibnamefont {Yang}}, \bibinfo {author} {\bibfnamefont
  {T.}~\bibnamefont {Xu}}, \bibinfo {author} {\bibfnamefont {B.}~\bibnamefont
  {Deng}}, \bibinfo {author} {\bibfnamefont {S.}~\bibnamefont {Stemmer}},
  \bibinfo {author} {\bibfnamefont {J.~A.}\ \bibnamefont {Schuller}}, \ and\
  \bibinfo {author} {\bibfnamefont {B.}~\bibnamefont {Liao}},\ }\href@noop {}
  {\bibfield  {journal} {\bibinfo  {journal} {Physical Review Research}\
  }\textbf {\bibinfo {volume} {1}},\ \bibinfo {pages} {033101} (\bibinfo {year}
  {2019}{\natexlab{a}})}\BibitemShut {NoStop}%
\bibitem [{\citenamefont {Kohn}(1959)}]{Kohn1959}%
  \BibitemOpen
  \bibfield  {author} {\bibinfo {author} {\bibfnamefont {W.}~\bibnamefont
  {Kohn}},\ }\href@noop {} {\bibfield  {journal} {\bibinfo  {journal} {Physical
  Review Letters}\ }\textbf {\bibinfo {volume} {2}},\ \bibinfo {pages} {393}
  (\bibinfo {year} {1959})}\BibitemShut {NoStop}%
\bibitem [{\citenamefont {Nguyen}\ \emph {et~al.}(2020)\citenamefont {Nguyen},
  \citenamefont {Han}, \citenamefont {Andrejevic}, \citenamefont {Pablo-Pedro},
  \citenamefont {Apte}, \citenamefont {Tsurimaki}, \citenamefont {Ding},
  \citenamefont {Zhang}, \citenamefont {Alatas}, \citenamefont {Alp} \emph
  {et~al.}}]{nguyen2020topological}%
  \BibitemOpen
  \bibfield  {author} {\bibinfo {author} {\bibfnamefont {T.}~\bibnamefont
  {Nguyen}}, \bibinfo {author} {\bibfnamefont {F.}~\bibnamefont {Han}},
  \bibinfo {author} {\bibfnamefont {N.}~\bibnamefont {Andrejevic}}, \bibinfo
  {author} {\bibfnamefont {R.}~\bibnamefont {Pablo-Pedro}}, \bibinfo {author}
  {\bibfnamefont {A.}~\bibnamefont {Apte}}, \bibinfo {author} {\bibfnamefont
  {Y.}~\bibnamefont {Tsurimaki}}, \bibinfo {author} {\bibfnamefont
  {Z.}~\bibnamefont {Ding}}, \bibinfo {author} {\bibfnamefont {K.}~\bibnamefont
  {Zhang}}, \bibinfo {author} {\bibfnamefont {A.}~\bibnamefont {Alatas}},
  \bibinfo {author} {\bibfnamefont {E.~E.}\ \bibnamefont {Alp}},  \emph
  {et~al.},\ }\href@noop {} {\bibfield  {journal} {\bibinfo  {journal}
  {Physical Review Letters}\ }\textbf {\bibinfo {volume} {124}},\ \bibinfo
  {pages} {236401} (\bibinfo {year} {2020})}\BibitemShut {NoStop}%
\bibitem [{\citenamefont {Weng}\ \emph {et~al.}(2014)\citenamefont {Weng},
  \citenamefont {Dai},\ and\ \citenamefont {Fang}}]{weng2014}%
  \BibitemOpen
  \bibfield  {author} {\bibinfo {author} {\bibfnamefont {H.}~\bibnamefont
  {Weng}}, \bibinfo {author} {\bibfnamefont {X.}~\bibnamefont {Dai}}, \ and\
  \bibinfo {author} {\bibfnamefont {Z.}~\bibnamefont {Fang}},\ }\href@noop {}
  {\bibfield  {journal} {\bibinfo  {journal} {Phys. Rev. X}\ }\textbf {\bibinfo
  {volume} {4}},\ \bibinfo {pages} {011002} (\bibinfo {year}
  {2014})}\BibitemShut {NoStop}%
\bibitem [{\citenamefont {Zhou}\ \emph {et~al.}(2016)\citenamefont {Zhou},
  \citenamefont {Wu}, \citenamefont {Ning}, \citenamefont {Li}, \citenamefont
  {Du}, \citenamefont {Chen}, \citenamefont {Zhang}, \citenamefont {Chi},
  \citenamefont {Wang}, \citenamefont {Zhu} \emph {et~al.}}]{zhou2016pressure}%
  \BibitemOpen
  \bibfield  {author} {\bibinfo {author} {\bibfnamefont {Y.}~\bibnamefont
  {Zhou}}, \bibinfo {author} {\bibfnamefont {J.}~\bibnamefont {Wu}}, \bibinfo
  {author} {\bibfnamefont {W.}~\bibnamefont {Ning}}, \bibinfo {author}
  {\bibfnamefont {N.}~\bibnamefont {Li}}, \bibinfo {author} {\bibfnamefont
  {Y.}~\bibnamefont {Du}}, \bibinfo {author} {\bibfnamefont {X.}~\bibnamefont
  {Chen}}, \bibinfo {author} {\bibfnamefont {R.}~\bibnamefont {Zhang}},
  \bibinfo {author} {\bibfnamefont {Z.}~\bibnamefont {Chi}}, \bibinfo {author}
  {\bibfnamefont {X.}~\bibnamefont {Wang}}, \bibinfo {author} {\bibfnamefont
  {X.}~\bibnamefont {Zhu}},  \emph {et~al.},\ }\href@noop {} {\bibfield
  {journal} {\bibinfo  {journal} {Proceedings of the National Academy of
  Sciences}\ }\textbf {\bibinfo {volume} {113}},\ \bibinfo {pages} {2904}
  (\bibinfo {year} {2016})}\BibitemShut {NoStop}%
\bibitem [{\citenamefont {Kresse}\ and\ \citenamefont
  {Furthm\"uller}(1996{\natexlab{a}})}]{vasp-01}%
  \BibitemOpen
  \bibfield  {author} {\bibinfo {author} {\bibfnamefont {G.}~\bibnamefont
  {Kresse}}\ and\ \bibinfo {author} {\bibfnamefont {J.}~\bibnamefont
  {Furthm\"uller}},\ }\href@noop {} {\bibfield  {journal} {\bibinfo  {journal}
  {Phys. Rev. B}\ }\textbf {\bibinfo {volume} {54}},\ \bibinfo {pages} {11169}
  (\bibinfo {year} {1996}{\natexlab{a}})}\BibitemShut {NoStop}%
\bibitem [{\citenamefont {Kresse}\ and\ \citenamefont
  {Furthm\"uller}(1996{\natexlab{b}})}]{vasp-02}%
  \BibitemOpen
  \bibfield  {author} {\bibinfo {author} {\bibfnamefont {G.}~\bibnamefont
  {Kresse}}\ and\ \bibinfo {author} {\bibfnamefont {J.}~\bibnamefont
  {Furthm\"uller}},\ }\href@noop {} {\bibfield  {journal} {\bibinfo  {journal}
  {Computational Materials Science}\ }\textbf {\bibinfo {volume} {6}},\
  \bibinfo {pages} {15} (\bibinfo {year} {1996}{\natexlab{b}})}\BibitemShut
  {NoStop}%
\bibitem [{\citenamefont {Bl\"ochl}(1994)}]{PAW-01}%
  \BibitemOpen
  \bibfield  {author} {\bibinfo {author} {\bibfnamefont {P.~E.}\ \bibnamefont
  {Bl\"ochl}},\ }\href@noop {} {\bibfield  {journal} {\bibinfo  {journal}
  {Phys. Rev. B}\ }\textbf {\bibinfo {volume} {50}},\ \bibinfo {pages} {17953}
  (\bibinfo {year} {1994})}\BibitemShut {NoStop}%
\bibitem [{\citenamefont {Kresse}\ and\ \citenamefont
  {Joubert}(1999)}]{PAW-02}%
  \BibitemOpen
  \bibfield  {author} {\bibinfo {author} {\bibfnamefont {G.}~\bibnamefont
  {Kresse}}\ and\ \bibinfo {author} {\bibfnamefont {D.}~\bibnamefont
  {Joubert}},\ }\href@noop {} {\bibfield  {journal} {\bibinfo  {journal} {Phys.
  Rev. B}\ }\textbf {\bibinfo {volume} {59}},\ \bibinfo {pages} {1758}
  (\bibinfo {year} {1999})}\BibitemShut {NoStop}%
\bibitem [{\citenamefont {Perdew}\ \emph {et~al.}(1996)\citenamefont {Perdew},
  \citenamefont {Burke},\ and\ \citenamefont {Ernzerhof}}]{GGA}%
  \BibitemOpen
  \bibfield  {author} {\bibinfo {author} {\bibfnamefont {J.~P.}\ \bibnamefont
  {Perdew}}, \bibinfo {author} {\bibfnamefont {K.}~\bibnamefont {Burke}}, \
  and\ \bibinfo {author} {\bibfnamefont {M.}~\bibnamefont {Ernzerhof}},\
  }\href@noop {} {\bibfield  {journal} {\bibinfo  {journal} {Phys. Rev. Lett.}\
  }\textbf {\bibinfo {volume} {77}},\ \bibinfo {pages} {3865} (\bibinfo {year}
  {1996})}\BibitemShut {NoStop}%
\bibitem [{\citenamefont {Klime{\v{s}}}\ \emph {et~al.}(2009)\citenamefont
  {Klime{\v{s}}}, \citenamefont {Bowler},\ and\ \citenamefont
  {Michaelides}}]{vdw-01}%
  \BibitemOpen
  \bibfield  {author} {\bibinfo {author} {\bibfnamefont {J.}~\bibnamefont
  {Klime{\v{s}}}}, \bibinfo {author} {\bibfnamefont {D.~R.}\ \bibnamefont
  {Bowler}}, \ and\ \bibinfo {author} {\bibfnamefont {A.}~\bibnamefont
  {Michaelides}},\ }\href@noop {} {\bibfield  {journal} {\bibinfo  {journal}
  {Journal of Physics: Condensed Matter}\ }\textbf {\bibinfo {volume} {22}},\
  \bibinfo {pages} {022201} (\bibinfo {year} {2009})}\BibitemShut {NoStop}%
\bibitem [{\citenamefont {Klime\ifmmode~\check{s}\else \v{s}\fi{}}\ \emph
  {et~al.}(2011)\citenamefont {Klime\ifmmode~\check{s}\else \v{s}\fi{}},
  \citenamefont {Bowler},\ and\ \citenamefont {Michaelides}}]{vdw-02}%
  \BibitemOpen
  \bibfield  {author} {\bibinfo {author} {\bibfnamefont {J.~c.~v.}\
  \bibnamefont {Klime\ifmmode~\check{s}\else \v{s}\fi{}}}, \bibinfo {author}
  {\bibfnamefont {D.~R.}\ \bibnamefont {Bowler}}, \ and\ \bibinfo {author}
  {\bibfnamefont {A.}~\bibnamefont {Michaelides}},\ }\href@noop {} {\bibfield
  {journal} {\bibinfo  {journal} {Phys. Rev. B}\ }\textbf {\bibinfo {volume}
  {83}},\ \bibinfo {pages} {195131} (\bibinfo {year} {2011})}\BibitemShut
  {NoStop}%
\bibitem [{\citenamefont {Fan}\ \emph {et~al.}(2017{\natexlab{b}})\citenamefont
  {Fan}, \citenamefont {Liang}, \citenamefont {Chen}, \citenamefont {Yao},\
  and\ \citenamefont {Zhou}}]{ZrTe5-01}%
  \BibitemOpen
  \bibfield  {author} {\bibinfo {author} {\bibfnamefont {Z.}~\bibnamefont
  {Fan}}, \bibinfo {author} {\bibfnamefont {Q.-F.}\ \bibnamefont {Liang}},
  \bibinfo {author} {\bibfnamefont {Y.}~\bibnamefont {Chen}}, \bibinfo {author}
  {\bibfnamefont {S.-H.}\ \bibnamefont {Yao}}, \ and\ \bibinfo {author}
  {\bibfnamefont {J.}~\bibnamefont {Zhou}},\ }\href@noop {} {\bibfield
  {journal} {\bibinfo  {journal} {Scientific Reports}\ }\textbf {\bibinfo
  {volume} {7}},\ \bibinfo {pages} {45667} (\bibinfo {year}
  {2017}{\natexlab{b}})}\BibitemShut {NoStop}%
\bibitem [{\citenamefont {Wang}\ \emph
  {et~al.}(2018{\natexlab{a}})\citenamefont {Wang}, \citenamefont {Wang},
  \citenamefont {Chen}, \citenamefont {Yao},\ and\ \citenamefont
  {Zhou}}]{ZrTe5-02}%
  \BibitemOpen
  \bibfield  {author} {\bibinfo {author} {\bibfnamefont {C.}~\bibnamefont
  {Wang}}, \bibinfo {author} {\bibfnamefont {H.}~\bibnamefont {Wang}}, \bibinfo
  {author} {\bibfnamefont {Y.}~\bibnamefont {Chen}}, \bibinfo {author}
  {\bibfnamefont {S.-H.}\ \bibnamefont {Yao}}, \ and\ \bibinfo {author}
  {\bibfnamefont {J.}~\bibnamefont {Zhou}},\ }\href@noop {} {\bibfield
  {journal} {\bibinfo  {journal} {Journal of Applied Physics}\ }\textbf
  {\bibinfo {volume} {123}},\ \bibinfo {pages} {175104} (\bibinfo {year}
  {2018}{\natexlab{a}})}\BibitemShut {NoStop}%
\bibitem [{\citenamefont {Togo}\ \emph {et~al.}(2008)\citenamefont {Togo},
  \citenamefont {Oba},\ and\ \citenamefont {Tanaka}}]{phonopy}%
  \BibitemOpen
  \bibfield  {author} {\bibinfo {author} {\bibfnamefont {A.}~\bibnamefont
  {Togo}}, \bibinfo {author} {\bibfnamefont {F.}~\bibnamefont {Oba}}, \ and\
  \bibinfo {author} {\bibfnamefont {I.}~\bibnamefont {Tanaka}},\ }\href@noop {}
  {\bibfield  {journal} {\bibinfo  {journal} {Phys. Rev. B}\ }\textbf {\bibinfo
  {volume} {78}},\ \bibinfo {pages} {134106} (\bibinfo {year}
  {2008})}\BibitemShut {NoStop}%
\bibitem [{\citenamefont {Wang}\ \emph
  {et~al.}(2018{\natexlab{b}})\citenamefont {Wang}, \citenamefont {Wang},
  \citenamefont {Chen}, \citenamefont {Yao},\ and\ \citenamefont
  {Zhou}}]{wang2018first}%
  \BibitemOpen
  \bibfield  {author} {\bibinfo {author} {\bibfnamefont {C.}~\bibnamefont
  {Wang}}, \bibinfo {author} {\bibfnamefont {H.}~\bibnamefont {Wang}}, \bibinfo
  {author} {\bibfnamefont {Y.}~\bibnamefont {Chen}}, \bibinfo {author}
  {\bibfnamefont {S.-H.}\ \bibnamefont {Yao}}, \ and\ \bibinfo {author}
  {\bibfnamefont {J.}~\bibnamefont {Zhou}},\ }\href@noop {} {\bibfield
  {journal} {\bibinfo  {journal} {Journal of Applied Physics}\ }\textbf
  {\bibinfo {volume} {123}},\ \bibinfo {pages} {175104} (\bibinfo {year}
  {2018}{\natexlab{b}})}\BibitemShut {NoStop}%
\bibitem [{\citenamefont {Li}\ \emph {et~al.}(2014)\citenamefont {Li},
  \citenamefont {Carrete}, \citenamefont {Katcho},\ and\ \citenamefont
  {Mingo}}]{shengbte-01}%
  \BibitemOpen
  \bibfield  {author} {\bibinfo {author} {\bibfnamefont {W.}~\bibnamefont
  {Li}}, \bibinfo {author} {\bibfnamefont {J.}~\bibnamefont {Carrete}},
  \bibinfo {author} {\bibfnamefont {N.~A.}\ \bibnamefont {Katcho}}, \ and\
  \bibinfo {author} {\bibfnamefont {N.}~\bibnamefont {Mingo}},\ }\href@noop {}
  {\bibfield  {journal} {\bibinfo  {journal} {Computer Physics Communications}\
  }\textbf {\bibinfo {volume} {185}},\ \bibinfo {pages} {1747 } (\bibinfo
  {year} {2014})}\BibitemShut {NoStop}%
\bibitem [{\citenamefont {Li}\ \emph {et~al.}(2012)\citenamefont {Li},
  \citenamefont {Mingo}, \citenamefont {Lindsay}, \citenamefont {Broido},
  \citenamefont {Stewart},\ and\ \citenamefont {Katcho}}]{shengbte-02}%
  \BibitemOpen
  \bibfield  {author} {\bibinfo {author} {\bibfnamefont {W.}~\bibnamefont
  {Li}}, \bibinfo {author} {\bibfnamefont {N.}~\bibnamefont {Mingo}}, \bibinfo
  {author} {\bibfnamefont {L.}~\bibnamefont {Lindsay}}, \bibinfo {author}
  {\bibfnamefont {D.~A.}\ \bibnamefont {Broido}}, \bibinfo {author}
  {\bibfnamefont {D.~A.}\ \bibnamefont {Stewart}}, \ and\ \bibinfo {author}
  {\bibfnamefont {N.~A.}\ \bibnamefont {Katcho}},\ }\href@noop {} {\bibfield
  {journal} {\bibinfo  {journal} {Phys. Rev. B}\ }\textbf {\bibinfo {volume}
  {85}},\ \bibinfo {pages} {195436} (\bibinfo {year} {2012})}\BibitemShut
  {NoStop}%
\bibitem [{\citenamefont {Piscanec}\ \emph {et~al.}(2004)\citenamefont
  {Piscanec}, \citenamefont {Lazzeri}, \citenamefont {Mauri}, \citenamefont
  {Ferrari},\ and\ \citenamefont {Robertson}}]{piscanec2004kohn}%
  \BibitemOpen
  \bibfield  {author} {\bibinfo {author} {\bibfnamefont {S.}~\bibnamefont
  {Piscanec}}, \bibinfo {author} {\bibfnamefont {M.}~\bibnamefont {Lazzeri}},
  \bibinfo {author} {\bibfnamefont {F.}~\bibnamefont {Mauri}}, \bibinfo
  {author} {\bibfnamefont {A.}~\bibnamefont {Ferrari}}, \ and\ \bibinfo
  {author} {\bibfnamefont {J.}~\bibnamefont {Robertson}},\ }\href@noop {}
  {\bibfield  {journal} {\bibinfo  {journal} {Physical Review Letters}\
  }\textbf {\bibinfo {volume} {93}},\ \bibinfo {pages} {185503} (\bibinfo
  {year} {2004})}\BibitemShut {NoStop}%
\bibitem [{\citenamefont {Lazzeri}\ and\ \citenamefont
  {Mauri}(2006)}]{lazzeri2006nonadiabatic}%
  \BibitemOpen
  \bibfield  {author} {\bibinfo {author} {\bibfnamefont {M.}~\bibnamefont
  {Lazzeri}}\ and\ \bibinfo {author} {\bibfnamefont {F.}~\bibnamefont
  {Mauri}},\ }\href@noop {} {\bibfield  {journal} {\bibinfo  {journal}
  {Physical Review Letters}\ }\textbf {\bibinfo {volume} {97}},\ \bibinfo
  {pages} {266407} (\bibinfo {year} {2006})}\BibitemShut {NoStop}%
\bibitem [{\citenamefont {Noffsinger}\ \emph {et~al.}(2010)\citenamefont
  {Noffsinger}, \citenamefont {Giustino}, \citenamefont {Malone}, \citenamefont
  {Park}, \citenamefont {Louie},\ and\ \citenamefont
  {Cohen}}]{noffsinger2010epw}%
  \BibitemOpen
  \bibfield  {author} {\bibinfo {author} {\bibfnamefont {J.}~\bibnamefont
  {Noffsinger}}, \bibinfo {author} {\bibfnamefont {F.}~\bibnamefont
  {Giustino}}, \bibinfo {author} {\bibfnamefont {B.~D.}\ \bibnamefont
  {Malone}}, \bibinfo {author} {\bibfnamefont {C.-H.}\ \bibnamefont {Park}},
  \bibinfo {author} {\bibfnamefont {S.~G.}\ \bibnamefont {Louie}}, \ and\
  \bibinfo {author} {\bibfnamefont {M.~L.}\ \bibnamefont {Cohen}},\ }\href@noop
  {} {\bibfield  {journal} {\bibinfo  {journal} {Computer Physics
  Communications}\ }\textbf {\bibinfo {volume} {181}},\ \bibinfo {pages} {2140}
  (\bibinfo {year} {2010})}\BibitemShut {NoStop}%
\bibitem [{\citenamefont {Giannozzi}\ \emph {et~al.}(2017)\citenamefont
  {Giannozzi}, \citenamefont {Andreussi}, \citenamefont {Brumme}, \citenamefont
  {Bunau}, \citenamefont {Nardelli}, \citenamefont {Calandra}, \citenamefont
  {Car}, \citenamefont {Cavazzoni}, \citenamefont {Ceresoli}, \citenamefont
  {Cococcioni}, \citenamefont {Colonna}, \citenamefont {Carnimeo},
  \citenamefont {Corso}, \citenamefont {de~Gironcoli}, \citenamefont {Delugas},
  \citenamefont {DiStasio}, \citenamefont {Ferretti}, \citenamefont {Floris},
  \citenamefont {Fratesi}, \citenamefont {Fugallo}, \citenamefont {Gebauer},
  \citenamefont {Gerstmann}, \citenamefont {Giustino}, \citenamefont {Gorni},
  \citenamefont {Jia}, \citenamefont {Kawamura}, \citenamefont {Ko},
  \citenamefont {Kokalj}, \citenamefont {Kü{\c{c}}ükbenli}, \citenamefont
  {Lazzeri}, \citenamefont {Marsili}, \citenamefont {Marzari}, \citenamefont
  {Mauri}, \citenamefont {Nguyen}, \citenamefont {Nguyen}, \citenamefont {de-la
  Roza}, \citenamefont {Paulatto}, \citenamefont {Ponc{\'{e}}}, \citenamefont
  {Rocca}, \citenamefont {Sabatini}, \citenamefont {Santra}, \citenamefont
  {Schlipf}, \citenamefont {Seitsonen}, \citenamefont {Smogunov}, \citenamefont
  {Timrov}, \citenamefont {Thonhauser}, \citenamefont {Umari}, \citenamefont
  {Vast}, \citenamefont {Wu},\ and\ \citenamefont {Baroni}}]{Giannozzi_2017}%
  \BibitemOpen
  \bibfield  {author} {\bibinfo {author} {\bibfnamefont {P.}~\bibnamefont
  {Giannozzi}}, \bibinfo {author} {\bibfnamefont {O.}~\bibnamefont
  {Andreussi}}, \bibinfo {author} {\bibfnamefont {T.}~\bibnamefont {Brumme}},
  \bibinfo {author} {\bibfnamefont {O.}~\bibnamefont {Bunau}}, \bibinfo
  {author} {\bibfnamefont {M.~B.}\ \bibnamefont {Nardelli}}, \bibinfo {author}
  {\bibfnamefont {M.}~\bibnamefont {Calandra}}, \bibinfo {author}
  {\bibfnamefont {R.}~\bibnamefont {Car}}, \bibinfo {author} {\bibfnamefont
  {C.}~\bibnamefont {Cavazzoni}}, \bibinfo {author} {\bibfnamefont
  {D.}~\bibnamefont {Ceresoli}}, \bibinfo {author} {\bibfnamefont
  {M.}~\bibnamefont {Cococcioni}}, \bibinfo {author} {\bibfnamefont
  {N.}~\bibnamefont {Colonna}}, \bibinfo {author} {\bibfnamefont
  {I.}~\bibnamefont {Carnimeo}}, \bibinfo {author} {\bibfnamefont {A.~D.}\
  \bibnamefont {Corso}}, \bibinfo {author} {\bibfnamefont {S.}~\bibnamefont
  {de~Gironcoli}}, \bibinfo {author} {\bibfnamefont {P.}~\bibnamefont
  {Delugas}}, \bibinfo {author} {\bibfnamefont {R.~A.}\ \bibnamefont
  {DiStasio}}, \bibinfo {author} {\bibfnamefont {A.}~\bibnamefont {Ferretti}},
  \bibinfo {author} {\bibfnamefont {A.}~\bibnamefont {Floris}}, \bibinfo
  {author} {\bibfnamefont {G.}~\bibnamefont {Fratesi}}, \bibinfo {author}
  {\bibfnamefont {G.}~\bibnamefont {Fugallo}}, \bibinfo {author} {\bibfnamefont
  {R.}~\bibnamefont {Gebauer}}, \bibinfo {author} {\bibfnamefont
  {U.}~\bibnamefont {Gerstmann}}, \bibinfo {author} {\bibfnamefont
  {F.}~\bibnamefont {Giustino}}, \bibinfo {author} {\bibfnamefont
  {T.}~\bibnamefont {Gorni}}, \bibinfo {author} {\bibfnamefont
  {J.}~\bibnamefont {Jia}}, \bibinfo {author} {\bibfnamefont {M.}~\bibnamefont
  {Kawamura}}, \bibinfo {author} {\bibfnamefont {H.-Y.}\ \bibnamefont {Ko}},
  \bibinfo {author} {\bibfnamefont {A.}~\bibnamefont {Kokalj}}, \bibinfo
  {author} {\bibfnamefont {E.}~\bibnamefont {Kü{\c{c}}ükbenli}}, \bibinfo
  {author} {\bibfnamefont {M.}~\bibnamefont {Lazzeri}}, \bibinfo {author}
  {\bibfnamefont {M.}~\bibnamefont {Marsili}}, \bibinfo {author} {\bibfnamefont
  {N.}~\bibnamefont {Marzari}}, \bibinfo {author} {\bibfnamefont
  {F.}~\bibnamefont {Mauri}}, \bibinfo {author} {\bibfnamefont {N.~L.}\
  \bibnamefont {Nguyen}}, \bibinfo {author} {\bibfnamefont {H.-V.}\
  \bibnamefont {Nguyen}}, \bibinfo {author} {\bibfnamefont {A.~O.}\
  \bibnamefont {de-la Roza}}, \bibinfo {author} {\bibfnamefont
  {L.}~\bibnamefont {Paulatto}}, \bibinfo {author} {\bibfnamefont
  {S.}~\bibnamefont {Ponc{\'{e}}}}, \bibinfo {author} {\bibfnamefont
  {D.}~\bibnamefont {Rocca}}, \bibinfo {author} {\bibfnamefont
  {R.}~\bibnamefont {Sabatini}}, \bibinfo {author} {\bibfnamefont
  {B.}~\bibnamefont {Santra}}, \bibinfo {author} {\bibfnamefont
  {M.}~\bibnamefont {Schlipf}}, \bibinfo {author} {\bibfnamefont {A.~P.}\
  \bibnamefont {Seitsonen}}, \bibinfo {author} {\bibfnamefont {A.}~\bibnamefont
  {Smogunov}}, \bibinfo {author} {\bibfnamefont {I.}~\bibnamefont {Timrov}},
  \bibinfo {author} {\bibfnamefont {T.}~\bibnamefont {Thonhauser}}, \bibinfo
  {author} {\bibfnamefont {P.}~\bibnamefont {Umari}}, \bibinfo {author}
  {\bibfnamefont {N.}~\bibnamefont {Vast}}, \bibinfo {author} {\bibfnamefont
  {X.}~\bibnamefont {Wu}}, \ and\ \bibinfo {author} {\bibfnamefont
  {S.}~\bibnamefont {Baroni}},\ }\href {\doibase 10.1088/1361-648x/aa8f79}
  {\bibfield  {journal} {\bibinfo  {journal} {Journal of Physics: Condensed
  Matter}\ }\textbf {\bibinfo {volume} {29}},\ \bibinfo {pages} {465901}
  (\bibinfo {year} {2017})}\BibitemShut {NoStop}%
\bibitem [{\citenamefont {Liu}\ \emph {et~al.}(2016)\citenamefont {Liu},
  \citenamefont {Yuan}, \citenamefont {Zhang}, \citenamefont {Jin},
  \citenamefont {Narayan}, \citenamefont {Luo}, \citenamefont {Chen},
  \citenamefont {Yang}, \citenamefont {Zou}, \citenamefont {Wu} \emph
  {et~al.}}]{liu2016zeeman}%
  \BibitemOpen
  \bibfield  {author} {\bibinfo {author} {\bibfnamefont {Y.}~\bibnamefont
  {Liu}}, \bibinfo {author} {\bibfnamefont {X.}~\bibnamefont {Yuan}}, \bibinfo
  {author} {\bibfnamefont {C.}~\bibnamefont {Zhang}}, \bibinfo {author}
  {\bibfnamefont {Z.}~\bibnamefont {Jin}}, \bibinfo {author} {\bibfnamefont
  {A.}~\bibnamefont {Narayan}}, \bibinfo {author} {\bibfnamefont
  {C.}~\bibnamefont {Luo}}, \bibinfo {author} {\bibfnamefont {Z.}~\bibnamefont
  {Chen}}, \bibinfo {author} {\bibfnamefont {L.}~\bibnamefont {Yang}}, \bibinfo
  {author} {\bibfnamefont {J.}~\bibnamefont {Zou}}, \bibinfo {author}
  {\bibfnamefont {X.}~\bibnamefont {Wu}},  \emph {et~al.},\ }\href@noop {}
  {\bibfield  {journal} {\bibinfo  {journal} {Nature Communications}\ }\textbf
  {\bibinfo {volume} {7}},\ \bibinfo {pages} {1} (\bibinfo {year}
  {2016})}\BibitemShut {NoStop}%
\bibitem [{\citenamefont {Nair}\ \emph {et~al.}(2018)\citenamefont {Nair},
  \citenamefont {Dumitrescu}, \citenamefont {Channa}, \citenamefont {Griffin},
  \citenamefont {Neaton}, \citenamefont {Potter},\ and\ \citenamefont
  {Analytis}}]{nair2018thermodynamic}%
  \BibitemOpen
  \bibfield  {author} {\bibinfo {author} {\bibfnamefont {N.~L.}\ \bibnamefont
  {Nair}}, \bibinfo {author} {\bibfnamefont {P.~T.}\ \bibnamefont
  {Dumitrescu}}, \bibinfo {author} {\bibfnamefont {S.}~\bibnamefont {Channa}},
  \bibinfo {author} {\bibfnamefont {S.~M.}\ \bibnamefont {Griffin}}, \bibinfo
  {author} {\bibfnamefont {J.~B.}\ \bibnamefont {Neaton}}, \bibinfo {author}
  {\bibfnamefont {A.~C.}\ \bibnamefont {Potter}}, \ and\ \bibinfo {author}
  {\bibfnamefont {J.~G.}\ \bibnamefont {Analytis}},\ }\href@noop {} {\bibfield
  {journal} {\bibinfo  {journal} {Physical Review B}\ }\textbf {\bibinfo
  {volume} {97}},\ \bibinfo {pages} {041111} (\bibinfo {year}
  {2018})}\BibitemShut {NoStop}%
\bibitem [{\citenamefont {Chen}\ \emph {et~al.}(2015)\citenamefont {Chen},
  \citenamefont {Zhang}, \citenamefont {Schneeloch}, \citenamefont {Zhang},
  \citenamefont {Li}, \citenamefont {Gu},\ and\ \citenamefont
  {Wang}}]{chen2015optical}%
  \BibitemOpen
  \bibfield  {author} {\bibinfo {author} {\bibfnamefont {R.}~\bibnamefont
  {Chen}}, \bibinfo {author} {\bibfnamefont {S.}~\bibnamefont {Zhang}},
  \bibinfo {author} {\bibfnamefont {J.}~\bibnamefont {Schneeloch}}, \bibinfo
  {author} {\bibfnamefont {C.}~\bibnamefont {Zhang}}, \bibinfo {author}
  {\bibfnamefont {Q.}~\bibnamefont {Li}}, \bibinfo {author} {\bibfnamefont
  {G.}~\bibnamefont {Gu}}, \ and\ \bibinfo {author} {\bibfnamefont
  {N.}~\bibnamefont {Wang}},\ }\href@noop {} {\bibfield  {journal} {\bibinfo
  {journal} {Physical Review B}\ }\textbf {\bibinfo {volume} {92}},\ \bibinfo
  {pages} {075107} (\bibinfo {year} {2015})}\BibitemShut {NoStop}%
\bibitem [{\citenamefont {Li}\ \emph {et~al.}(2016{\natexlab{a}})\citenamefont
  {Li}, \citenamefont {Kharzeev}, \citenamefont {Zhang}, \citenamefont {Huang},
  \citenamefont {Pletikosi{\'c}}, \citenamefont {Fedorov}, \citenamefont
  {Zhong}, \citenamefont {Schneeloch}, \citenamefont {Gu},\ and\ \citenamefont
  {Valla}}]{li2016chiral}%
  \BibitemOpen
  \bibfield  {author} {\bibinfo {author} {\bibfnamefont {Q.}~\bibnamefont
  {Li}}, \bibinfo {author} {\bibfnamefont {D.~E.}\ \bibnamefont {Kharzeev}},
  \bibinfo {author} {\bibfnamefont {C.}~\bibnamefont {Zhang}}, \bibinfo
  {author} {\bibfnamefont {Y.}~\bibnamefont {Huang}}, \bibinfo {author}
  {\bibfnamefont {I.}~\bibnamefont {Pletikosi{\'c}}}, \bibinfo {author}
  {\bibfnamefont {A.}~\bibnamefont {Fedorov}}, \bibinfo {author} {\bibfnamefont
  {R.}~\bibnamefont {Zhong}}, \bibinfo {author} {\bibfnamefont
  {J.}~\bibnamefont {Schneeloch}}, \bibinfo {author} {\bibfnamefont
  {G.}~\bibnamefont {Gu}}, \ and\ \bibinfo {author} {\bibfnamefont
  {T.}~\bibnamefont {Valla}},\ }\href@noop {} {\bibfield  {journal} {\bibinfo
  {journal} {Nature Physics}\ }\textbf {\bibinfo {volume} {12}},\ \bibinfo
  {pages} {550} (\bibinfo {year} {2016}{\natexlab{a}})}\BibitemShut {NoStop}%
\bibitem [{\citenamefont {Li}\ \emph {et~al.}(2016{\natexlab{b}})\citenamefont
  {Li}, \citenamefont {Huang}, \citenamefont {Lv}, \citenamefont {Zhang},
  \citenamefont {Yang}, \citenamefont {Zhang}, \citenamefont {Chen},
  \citenamefont {Yao}, \citenamefont {Zhou}, \citenamefont {Lu} \emph
  {et~al.}}]{li2016experimental}%
  \BibitemOpen
  \bibfield  {author} {\bibinfo {author} {\bibfnamefont {X.-B.}\ \bibnamefont
  {Li}}, \bibinfo {author} {\bibfnamefont {W.-K.}\ \bibnamefont {Huang}},
  \bibinfo {author} {\bibfnamefont {Y.-Y.}\ \bibnamefont {Lv}}, \bibinfo
  {author} {\bibfnamefont {K.-W.}\ \bibnamefont {Zhang}}, \bibinfo {author}
  {\bibfnamefont {C.-L.}\ \bibnamefont {Yang}}, \bibinfo {author}
  {\bibfnamefont {B.-B.}\ \bibnamefont {Zhang}}, \bibinfo {author}
  {\bibfnamefont {Y.}~\bibnamefont {Chen}}, \bibinfo {author} {\bibfnamefont
  {S.-H.}\ \bibnamefont {Yao}}, \bibinfo {author} {\bibfnamefont
  {J.}~\bibnamefont {Zhou}}, \bibinfo {author} {\bibfnamefont {M.-H.}\
  \bibnamefont {Lu}},  \emph {et~al.},\ }\href@noop {} {\bibfield  {journal}
  {\bibinfo  {journal} {Physical Review Letters}\ }\textbf {\bibinfo {volume}
  {116}},\ \bibinfo {pages} {176803} (\bibinfo {year}
  {2016}{\natexlab{b}})}\BibitemShut {NoStop}%
\bibitem [{\citenamefont {Chen}\ \emph {et~al.}(2017)\citenamefont {Chen},
  \citenamefont {Chen}, \citenamefont {Zhong}, \citenamefont {Schneeloch},
  \citenamefont {Zhang}, \citenamefont {Huang}, \citenamefont {Qu},
  \citenamefont {Yu}, \citenamefont {Li}, \citenamefont {Gu} \emph
  {et~al.}}]{chen2017spectroscopic}%
  \BibitemOpen
  \bibfield  {author} {\bibinfo {author} {\bibfnamefont {Z.-G.}\ \bibnamefont
  {Chen}}, \bibinfo {author} {\bibfnamefont {R.}~\bibnamefont {Chen}}, \bibinfo
  {author} {\bibfnamefont {R.}~\bibnamefont {Zhong}}, \bibinfo {author}
  {\bibfnamefont {J.}~\bibnamefont {Schneeloch}}, \bibinfo {author}
  {\bibfnamefont {C.}~\bibnamefont {Zhang}}, \bibinfo {author} {\bibfnamefont
  {Y.}~\bibnamefont {Huang}}, \bibinfo {author} {\bibfnamefont
  {F.}~\bibnamefont {Qu}}, \bibinfo {author} {\bibfnamefont {R.}~\bibnamefont
  {Yu}}, \bibinfo {author} {\bibfnamefont {Q.}~\bibnamefont {Li}}, \bibinfo
  {author} {\bibfnamefont {G.}~\bibnamefont {Gu}},  \emph {et~al.},\
  }\href@noop {} {\bibfield  {journal} {\bibinfo  {journal} {Proceedings of the
  National Academy of Sciences}\ }\textbf {\bibinfo {volume} {114}},\ \bibinfo
  {pages} {816} (\bibinfo {year} {2017})}\BibitemShut {NoStop}%
\bibitem [{\citenamefont {Jiang}\ \emph {et~al.}(2017)\citenamefont {Jiang},
  \citenamefont {Dun}, \citenamefont {Zhou}, \citenamefont {Lu}, \citenamefont
  {Chen}, \citenamefont {Moon}, \citenamefont {Besara}, \citenamefont
  {Siegrist}, \citenamefont {Baumbach}, \citenamefont {Smirnov} \emph
  {et~al.}}]{jiang2017landau}%
  \BibitemOpen
  \bibfield  {author} {\bibinfo {author} {\bibfnamefont {Y.}~\bibnamefont
  {Jiang}}, \bibinfo {author} {\bibfnamefont {Z.}~\bibnamefont {Dun}}, \bibinfo
  {author} {\bibfnamefont {H.}~\bibnamefont {Zhou}}, \bibinfo {author}
  {\bibfnamefont {Z.}~\bibnamefont {Lu}}, \bibinfo {author} {\bibfnamefont
  {K.-W.}\ \bibnamefont {Chen}}, \bibinfo {author} {\bibfnamefont
  {S.}~\bibnamefont {Moon}}, \bibinfo {author} {\bibfnamefont {T.}~\bibnamefont
  {Besara}}, \bibinfo {author} {\bibfnamefont {T.}~\bibnamefont {Siegrist}},
  \bibinfo {author} {\bibfnamefont {R.}~\bibnamefont {Baumbach}}, \bibinfo
  {author} {\bibfnamefont {D.}~\bibnamefont {Smirnov}},  \emph {et~al.},\
  }\href@noop {} {\bibfield  {journal} {\bibinfo  {journal} {Physical Review
  B}\ }\textbf {\bibinfo {volume} {96}},\ \bibinfo {pages} {041101} (\bibinfo
  {year} {2017})}\BibitemShut {NoStop}%
\bibitem [{\citenamefont {Xiong}\ \emph {et~al.}(2017)\citenamefont {Xiong},
  \citenamefont {Sobota}, \citenamefont {Yang}, \citenamefont {Soifer},
  \citenamefont {Gauthier}, \citenamefont {Lu}, \citenamefont {Lv},
  \citenamefont {Yao}, \citenamefont {Lu}, \citenamefont {Hashimoto} \emph
  {et~al.}}]{xiong2017three}%
  \BibitemOpen
  \bibfield  {author} {\bibinfo {author} {\bibfnamefont {H.}~\bibnamefont
  {Xiong}}, \bibinfo {author} {\bibfnamefont {J.}~\bibnamefont {Sobota}},
  \bibinfo {author} {\bibfnamefont {S.-L.}\ \bibnamefont {Yang}}, \bibinfo
  {author} {\bibfnamefont {H.}~\bibnamefont {Soifer}}, \bibinfo {author}
  {\bibfnamefont {A.}~\bibnamefont {Gauthier}}, \bibinfo {author}
  {\bibfnamefont {M.-H.}\ \bibnamefont {Lu}}, \bibinfo {author} {\bibfnamefont
  {Y.-Y.}\ \bibnamefont {Lv}}, \bibinfo {author} {\bibfnamefont {S.-H.}\
  \bibnamefont {Yao}}, \bibinfo {author} {\bibfnamefont {D.}~\bibnamefont
  {Lu}}, \bibinfo {author} {\bibfnamefont {M.}~\bibnamefont {Hashimoto}},
  \emph {et~al.},\ }\href@noop {} {\bibfield  {journal} {\bibinfo  {journal}
  {Physical Review B}\ }\textbf {\bibinfo {volume} {95}},\ \bibinfo {pages}
  {195119} (\bibinfo {year} {2017})}\BibitemShut {NoStop}%
\bibitem [{\citenamefont {Tang}\ \emph {et~al.}(2019)\citenamefont {Tang},
  \citenamefont {Ren}, \citenamefont {Wang}, \citenamefont {Zhong},
  \citenamefont {Schneeloch}, \citenamefont {Yang}, \citenamefont {Yang},
  \citenamefont {Lee}, \citenamefont {Gu}, \citenamefont {Qiao} \emph
  {et~al.}}]{tang2019three}%
  \BibitemOpen
  \bibfield  {author} {\bibinfo {author} {\bibfnamefont {F.}~\bibnamefont
  {Tang}}, \bibinfo {author} {\bibfnamefont {Y.}~\bibnamefont {Ren}}, \bibinfo
  {author} {\bibfnamefont {P.}~\bibnamefont {Wang}}, \bibinfo {author}
  {\bibfnamefont {R.}~\bibnamefont {Zhong}}, \bibinfo {author} {\bibfnamefont
  {J.}~\bibnamefont {Schneeloch}}, \bibinfo {author} {\bibfnamefont {S.~A.}\
  \bibnamefont {Yang}}, \bibinfo {author} {\bibfnamefont {K.}~\bibnamefont
  {Yang}}, \bibinfo {author} {\bibfnamefont {P.~A.}\ \bibnamefont {Lee}},
  \bibinfo {author} {\bibfnamefont {G.}~\bibnamefont {Gu}}, \bibinfo {author}
  {\bibfnamefont {Z.}~\bibnamefont {Qiao}},  \emph {et~al.},\ }\href@noop {}
  {\bibfield  {journal} {\bibinfo  {journal} {Nature}\ }\textbf {\bibinfo
  {volume} {569}},\ \bibinfo {pages} {537} (\bibinfo {year}
  {2019})}\BibitemShut {NoStop}%
\bibitem [{\citenamefont {Manzoni}\ \emph {et~al.}(2016)\citenamefont
  {Manzoni}, \citenamefont {Gragnaniello}, \citenamefont {Aut{\`e}s},
  \citenamefont {Kuhn}, \citenamefont {Sterzi}, \citenamefont {Cilento},
  \citenamefont {Zacchigna}, \citenamefont {Enenkel}, \citenamefont {Vobornik},
  \citenamefont {Barba} \emph {et~al.}}]{manzoni2016evidence}%
  \BibitemOpen
  \bibfield  {author} {\bibinfo {author} {\bibfnamefont {G.}~\bibnamefont
  {Manzoni}}, \bibinfo {author} {\bibfnamefont {L.}~\bibnamefont
  {Gragnaniello}}, \bibinfo {author} {\bibfnamefont {G.}~\bibnamefont
  {Aut{\`e}s}}, \bibinfo {author} {\bibfnamefont {T.}~\bibnamefont {Kuhn}},
  \bibinfo {author} {\bibfnamefont {A.}~\bibnamefont {Sterzi}}, \bibinfo
  {author} {\bibfnamefont {F.}~\bibnamefont {Cilento}}, \bibinfo {author}
  {\bibfnamefont {M.}~\bibnamefont {Zacchigna}}, \bibinfo {author}
  {\bibfnamefont {V.}~\bibnamefont {Enenkel}}, \bibinfo {author} {\bibfnamefont
  {I.}~\bibnamefont {Vobornik}}, \bibinfo {author} {\bibfnamefont
  {L.}~\bibnamefont {Barba}},  \emph {et~al.},\ }\href@noop {} {\bibfield
  {journal} {\bibinfo  {journal} {Physical Review Letters}\ }\textbf {\bibinfo
  {volume} {117}},\ \bibinfo {pages} {237601} (\bibinfo {year}
  {2016})}\BibitemShut {NoStop}%
\bibitem [{\citenamefont {Santos-Cottin}\ \emph {et~al.}(2020)\citenamefont
  {Santos-Cottin}, \citenamefont {Padlewski}, \citenamefont {Martino},
  \citenamefont {David}, \citenamefont {Le~Mardel{\'e}}, \citenamefont
  {Capitani}, \citenamefont {Borondics}, \citenamefont {Bachmann},
  \citenamefont {Putzke}, \citenamefont {Moll} \emph
  {et~al.}}]{santos2020probing}%
  \BibitemOpen
  \bibfield  {author} {\bibinfo {author} {\bibfnamefont {D.}~\bibnamefont
  {Santos-Cottin}}, \bibinfo {author} {\bibfnamefont {M.}~\bibnamefont
  {Padlewski}}, \bibinfo {author} {\bibfnamefont {E.}~\bibnamefont {Martino}},
  \bibinfo {author} {\bibfnamefont {S.~B.}\ \bibnamefont {David}}, \bibinfo
  {author} {\bibfnamefont {F.}~\bibnamefont {Le~Mardel{\'e}}}, \bibinfo
  {author} {\bibfnamefont {F.}~\bibnamefont {Capitani}}, \bibinfo {author}
  {\bibfnamefont {F.}~\bibnamefont {Borondics}}, \bibinfo {author}
  {\bibfnamefont {M.}~\bibnamefont {Bachmann}}, \bibinfo {author}
  {\bibfnamefont {C.}~\bibnamefont {Putzke}}, \bibinfo {author} {\bibfnamefont
  {P.}~\bibnamefont {Moll}},  \emph {et~al.},\ }\href@noop {} {\bibfield
  {journal} {\bibinfo  {journal} {Physical Review B}\ }\textbf {\bibinfo
  {volume} {101}},\ \bibinfo {pages} {125205} (\bibinfo {year}
  {2020})}\BibitemShut {NoStop}%
\bibitem [{\citenamefont {Martino}\ \emph {et~al.}(2019)\citenamefont
  {Martino}, \citenamefont {Crassee}, \citenamefont {Eguchi}, \citenamefont
  {Santos-Cottin}, \citenamefont {Zhong}, \citenamefont {Gu}, \citenamefont
  {Berger}, \citenamefont {Rukelj}, \citenamefont {Orlita}, \citenamefont
  {Homes} \emph {et~al.}}]{martino2019two}%
  \BibitemOpen
  \bibfield  {author} {\bibinfo {author} {\bibfnamefont {E.}~\bibnamefont
  {Martino}}, \bibinfo {author} {\bibfnamefont {I.}~\bibnamefont {Crassee}},
  \bibinfo {author} {\bibfnamefont {G.}~\bibnamefont {Eguchi}}, \bibinfo
  {author} {\bibfnamefont {D.}~\bibnamefont {Santos-Cottin}}, \bibinfo {author}
  {\bibfnamefont {R.}~\bibnamefont {Zhong}}, \bibinfo {author} {\bibfnamefont
  {G.}~\bibnamefont {Gu}}, \bibinfo {author} {\bibfnamefont {H.}~\bibnamefont
  {Berger}}, \bibinfo {author} {\bibfnamefont {Z.}~\bibnamefont {Rukelj}},
  \bibinfo {author} {\bibfnamefont {M.}~\bibnamefont {Orlita}}, \bibinfo
  {author} {\bibfnamefont {C.~C.}\ \bibnamefont {Homes}},  \emph {et~al.},\
  }\href@noop {} {\bibfield  {journal} {\bibinfo  {journal} {Physical Review
  Letters}\ }\textbf {\bibinfo {volume} {122}},\ \bibinfo {pages} {217402}
  (\bibinfo {year} {2019})}\BibitemShut {NoStop}%
\bibitem [{\citenamefont {Zhu}\ \emph {et~al.}(2018)\citenamefont {Zhu},
  \citenamefont {Feng}, \citenamefont {Mills}, \citenamefont {Wang},
  \citenamefont {Wu}, \citenamefont {Zhang}, \citenamefont {Pantelides},
  \citenamefont {Du},\ and\ \citenamefont {Wang}}]{zhu2018record}%
  \BibitemOpen
  \bibfield  {author} {\bibinfo {author} {\bibfnamefont {J.}~\bibnamefont
  {Zhu}}, \bibinfo {author} {\bibfnamefont {T.}~\bibnamefont {Feng}}, \bibinfo
  {author} {\bibfnamefont {S.}~\bibnamefont {Mills}}, \bibinfo {author}
  {\bibfnamefont {P.}~\bibnamefont {Wang}}, \bibinfo {author} {\bibfnamefont
  {X.}~\bibnamefont {Wu}}, \bibinfo {author} {\bibfnamefont {L.}~\bibnamefont
  {Zhang}}, \bibinfo {author} {\bibfnamefont {S.~T.}\ \bibnamefont
  {Pantelides}}, \bibinfo {author} {\bibfnamefont {X.}~\bibnamefont {Du}}, \
  and\ \bibinfo {author} {\bibfnamefont {X.}~\bibnamefont {Wang}},\ }\href@noop
  {} {\bibfield  {journal} {\bibinfo  {journal} {ACS Applied Materials \&
  Interfaces}\ }\textbf {\bibinfo {volume} {10}},\ \bibinfo {pages} {40740}
  (\bibinfo {year} {2018})}\BibitemShut {NoStop}%
\bibitem [{\citenamefont {Zaheer}\ \emph {et~al.}(2013)\citenamefont {Zaheer},
  \citenamefont {Young}, \citenamefont {Cellucci}, \citenamefont {Teo},
  \citenamefont {Kane}, \citenamefont {Mele},\ and\ \citenamefont
  {Rappe}}]{zaheer2013spin}%
  \BibitemOpen
  \bibfield  {author} {\bibinfo {author} {\bibfnamefont {S.}~\bibnamefont
  {Zaheer}}, \bibinfo {author} {\bibfnamefont {S.~M.}\ \bibnamefont {Young}},
  \bibinfo {author} {\bibfnamefont {D.}~\bibnamefont {Cellucci}}, \bibinfo
  {author} {\bibfnamefont {J.~C.}\ \bibnamefont {Teo}}, \bibinfo {author}
  {\bibfnamefont {C.~L.}\ \bibnamefont {Kane}}, \bibinfo {author}
  {\bibfnamefont {E.~J.}\ \bibnamefont {Mele}}, \ and\ \bibinfo {author}
  {\bibfnamefont {A.~M.}\ \bibnamefont {Rappe}},\ }\href@noop {} {\bibfield
  {journal} {\bibinfo  {journal} {Physical Review B}\ }\textbf {\bibinfo
  {volume} {87}},\ \bibinfo {pages} {045202} (\bibinfo {year}
  {2013})}\BibitemShut {NoStop}%
\bibitem [{\citenamefont {Rogalski}(2005)}]{rogalski2005hgcdte}%
  \BibitemOpen
  \bibfield  {author} {\bibinfo {author} {\bibfnamefont {A.}~\bibnamefont
  {Rogalski}},\ }\href@noop {} {\bibfield  {journal} {\bibinfo  {journal}
  {Reports on Progress in Physics}\ }\textbf {\bibinfo {volume} {68}},\
  \bibinfo {pages} {2267} (\bibinfo {year} {2005})}\BibitemShut {NoStop}%
\bibitem [{\citenamefont {Lee}\ \emph {et~al.}(2014)\citenamefont {Lee},
  \citenamefont {Esfarjani}, \citenamefont {Luo}, \citenamefont {Zhou},
  \citenamefont {Tian},\ and\ \citenamefont {Chen}}]{lee2014resonant}%
  \BibitemOpen
  \bibfield  {author} {\bibinfo {author} {\bibfnamefont {S.}~\bibnamefont
  {Lee}}, \bibinfo {author} {\bibfnamefont {K.}~\bibnamefont {Esfarjani}},
  \bibinfo {author} {\bibfnamefont {T.}~\bibnamefont {Luo}}, \bibinfo {author}
  {\bibfnamefont {J.}~\bibnamefont {Zhou}}, \bibinfo {author} {\bibfnamefont
  {Z.}~\bibnamefont {Tian}}, \ and\ \bibinfo {author} {\bibfnamefont
  {G.}~\bibnamefont {Chen}},\ }\href@noop {} {\bibfield  {journal} {\bibinfo
  {journal} {Nature Communications}\ }\textbf {\bibinfo {volume} {5}},\
  \bibinfo {pages} {1} (\bibinfo {year} {2014})}\BibitemShut {NoStop}%
\bibitem [{\citenamefont {Sheregii}\ \emph {et~al.}(2009)\citenamefont
  {Sheregii}, \citenamefont {Cebulski}, \citenamefont {Marcelli},\ and\
  \citenamefont {Piccinini}}]{sheregii2009temperature}%
  \BibitemOpen
  \bibfield  {author} {\bibinfo {author} {\bibfnamefont {E.}~\bibnamefont
  {Sheregii}}, \bibinfo {author} {\bibfnamefont {J.}~\bibnamefont {Cebulski}},
  \bibinfo {author} {\bibfnamefont {A.}~\bibnamefont {Marcelli}}, \ and\
  \bibinfo {author} {\bibfnamefont {M.}~\bibnamefont {Piccinini}},\ }\href@noop
  {} {\bibfield  {journal} {\bibinfo  {journal} {Physical Review Letters}\
  }\textbf {\bibinfo {volume} {102}},\ \bibinfo {pages} {045504} (\bibinfo
  {year} {2009})}\BibitemShut {NoStop}%
\bibitem [{\citenamefont {Wo{\'z}ny}\ \emph {et~al.}(2020)\citenamefont
  {Wo{\'z}ny}, \citenamefont {Szuszkiewicz}, \citenamefont {Dyksik},
  \citenamefont {Motyka}, \citenamefont {Szczerbakow}, \citenamefont
  {Bardyszewski}, \citenamefont {Story},\ and\ \citenamefont
  {Cebulski}}]{wozny2020electron}%
  \BibitemOpen
  \bibfield  {author} {\bibinfo {author} {\bibfnamefont {M.}~\bibnamefont
  {Wo{\'z}ny}}, \bibinfo {author} {\bibfnamefont {W.}~\bibnamefont
  {Szuszkiewicz}}, \bibinfo {author} {\bibfnamefont {M.}~\bibnamefont
  {Dyksik}}, \bibinfo {author} {\bibfnamefont {M.}~\bibnamefont {Motyka}},
  \bibinfo {author} {\bibfnamefont {A.}~\bibnamefont {Szczerbakow}}, \bibinfo
  {author} {\bibfnamefont {W.}~\bibnamefont {Bardyszewski}}, \bibinfo {author}
  {\bibfnamefont {T.}~\bibnamefont {Story}}, \ and\ \bibinfo {author}
  {\bibfnamefont {J.}~\bibnamefont {Cebulski}},\ }\href@noop {} {\bibfield
  {journal} {\bibinfo  {journal} {arXiv preprint arXiv:2003.11849}\ } (\bibinfo
  {year} {2020})}\BibitemShut {NoStop}%
\bibitem [{\citenamefont {Martyniuk}\ \emph {et~al.}(2010)\citenamefont
  {Martyniuk}, \citenamefont {Dell},\ and\ \citenamefont
  {Faraone}}]{martyniuk2010mechanical}%
  \BibitemOpen
  \bibfield  {author} {\bibinfo {author} {\bibfnamefont {M.}~\bibnamefont
  {Martyniuk}}, \bibinfo {author} {\bibfnamefont {J.}~\bibnamefont {Dell}}, \
  and\ \bibinfo {author} {\bibfnamefont {L.}~\bibnamefont {Faraone}},\
  }\href@noop {} {\bibfield  {journal} {\bibinfo  {journal} {Mercury Cadmium
  Telluride: Growth, Properties and Applications}\ ,\ \bibinfo {pages} {151}}
  (\bibinfo {year} {2010})}\BibitemShut {NoStop}%
\bibitem [{\citenamefont {Markov}\ \emph {et~al.}(2018)\citenamefont {Markov},
  \citenamefont {Hu}, \citenamefont {Liu}, \citenamefont {Liu}, \citenamefont
  {Poon}, \citenamefont {Esfarjani},\ and\ \citenamefont
  {Zebarjadi}}]{markov2018semi}%
  \BibitemOpen
  \bibfield  {author} {\bibinfo {author} {\bibfnamefont {M.}~\bibnamefont
  {Markov}}, \bibinfo {author} {\bibfnamefont {X.}~\bibnamefont {Hu}}, \bibinfo
  {author} {\bibfnamefont {H.-C.}\ \bibnamefont {Liu}}, \bibinfo {author}
  {\bibfnamefont {N.}~\bibnamefont {Liu}}, \bibinfo {author} {\bibfnamefont
  {S.~J.}\ \bibnamefont {Poon}}, \bibinfo {author} {\bibfnamefont
  {K.}~\bibnamefont {Esfarjani}}, \ and\ \bibinfo {author} {\bibfnamefont
  {M.}~\bibnamefont {Zebarjadi}},\ }\href@noop {} {\bibfield  {journal}
  {\bibinfo  {journal} {Scientific Reports}\ }\textbf {\bibinfo {volume} {8}},\
  \bibinfo {pages} {1} (\bibinfo {year} {2018})}\BibitemShut {NoStop}%
\bibitem [{\citenamefont {Liao}\ \emph {et~al.}(2015)\citenamefont {Liao},
  \citenamefont {Qiu}, \citenamefont {Zhou}, \citenamefont {Huberman},
  \citenamefont {Esfarjani},\ and\ \citenamefont {Chen}}]{liao2015significant}%
  \BibitemOpen
  \bibfield  {author} {\bibinfo {author} {\bibfnamefont {B.}~\bibnamefont
  {Liao}}, \bibinfo {author} {\bibfnamefont {B.}~\bibnamefont {Qiu}}, \bibinfo
  {author} {\bibfnamefont {J.}~\bibnamefont {Zhou}}, \bibinfo {author}
  {\bibfnamefont {S.}~\bibnamefont {Huberman}}, \bibinfo {author}
  {\bibfnamefont {K.}~\bibnamefont {Esfarjani}}, \ and\ \bibinfo {author}
  {\bibfnamefont {G.}~\bibnamefont {Chen}},\ }\href@noop {} {\bibfield
  {journal} {\bibinfo  {journal} {Physical Review Letters}\ }\textbf {\bibinfo
  {volume} {114}},\ \bibinfo {pages} {115901} (\bibinfo {year}
  {2015})}\BibitemShut {NoStop}%
\bibitem [{\citenamefont {Yue}\ \emph {et~al.}(2019{\natexlab{b}})\citenamefont
  {Yue}, \citenamefont {Yang},\ and\ \citenamefont
  {Liao}}]{yue2019controlling}%
  \BibitemOpen
  \bibfield  {author} {\bibinfo {author} {\bibfnamefont {S.-Y.}\ \bibnamefont
  {Yue}}, \bibinfo {author} {\bibfnamefont {R.}~\bibnamefont {Yang}}, \ and\
  \bibinfo {author} {\bibfnamefont {B.}~\bibnamefont {Liao}},\ }\href@noop {}
  {\bibfield  {journal} {\bibinfo  {journal} {Physical Review B}\ }\textbf
  {\bibinfo {volume} {100}},\ \bibinfo {pages} {115408} (\bibinfo {year}
  {2019}{\natexlab{b}})}\BibitemShut {NoStop}%
\bibitem [{\citenamefont {Wo{\'z}ny}\ \emph {et~al.}(2015)\citenamefont
  {Wo{\'z}ny}, \citenamefont {Cebulski}, \citenamefont {Marcelli},
  \citenamefont {Piccinini},\ and\ \citenamefont
  {Sheregii}}]{wozny2015influence}%
  \BibitemOpen
  \bibfield  {author} {\bibinfo {author} {\bibfnamefont {M.}~\bibnamefont
  {Wo{\'z}ny}}, \bibinfo {author} {\bibfnamefont {J.}~\bibnamefont {Cebulski}},
  \bibinfo {author} {\bibfnamefont {A.}~\bibnamefont {Marcelli}}, \bibinfo
  {author} {\bibfnamefont {M.}~\bibnamefont {Piccinini}}, \ and\ \bibinfo
  {author} {\bibfnamefont {E.}~\bibnamefont {Sheregii}},\ }\href@noop {}
  {\bibfield  {journal} {\bibinfo  {journal} {Journal of Applied Physics}\
  }\textbf {\bibinfo {volume} {117}},\ \bibinfo {pages} {025702} (\bibinfo
  {year} {2015})}\BibitemShut {NoStop}%
\bibitem [{\citenamefont {Kawamura}\ \emph {et~al.}(1974)\citenamefont
  {Kawamura}, \citenamefont {Katayama}, \citenamefont {Takano},\ and\
  \citenamefont {Hotta}}]{kawamura1974dielectric}%
  \BibitemOpen
  \bibfield  {author} {\bibinfo {author} {\bibfnamefont {H.}~\bibnamefont
  {Kawamura}}, \bibinfo {author} {\bibfnamefont {S.}~\bibnamefont {Katayama}},
  \bibinfo {author} {\bibfnamefont {S.}~\bibnamefont {Takano}}, \ and\ \bibinfo
  {author} {\bibfnamefont {S.}~\bibnamefont {Hotta}},\ }\href@noop {}
  {\bibfield  {journal} {\bibinfo  {journal} {Solid State Communications}\
  }\textbf {\bibinfo {volume} {14}},\ \bibinfo {pages} {259} (\bibinfo {year}
  {1974})}\BibitemShut {NoStop}%
\end{thebibliography}%

\end{document}